\documentclass[fleqn,usenatbib]{mnras}
\usepackage{newtxtext,newtxmath}
\usepackage{longtable}

\usepackage[T1]{fontenc}
\usepackage{ae,aecompl}

\usepackage{graphicx}	
\usepackage{amsmath}	
\usepackage{multirow}
\usepackage{booktabs}
\newcommand{\hii}{\mbox{H{\sc ii}~}}
\usepackage[colorinlistoftodos]{todonotes} 
\usepackage{threeparttable}

\title[Environmental effects on the IMF]{Testing the role of environmental effects on the initial mass function of low-mass stars}

\author[Belinda]{
Belinda Damian$^{1}$\thanks{belin\_93@yahoo.in},
Jessy Jose$^{1}$\thanks{jessyvjose1@gmail.com},  
Manash R. Samal$^2$, Estelle Moraux$^3$, Swagat R. Das$^1$, 
\newauthor
Sudeshna Patra$^1$
\\
$^{1}$Indian Institute of Science Education and Research (IISER) Tirupati, Rami Reddy Nagar, Karakambadi Road, Mangalam (P.O.), Tirupati 517 507, India\\
$^2$ Physical Research Laboratory (PRL), Navrangpura, Ahmedabad 380 009, Gujarat, India\\
$^3$ Univ. Grenoble Alpes, CNRS, IPAG, 38000 Grenoble, France\\
}

\date{Accepted XXX. Received YYY; in original form ZZZ}

\pubyear{2019}

\begin{document}
\label{firstpage}
\pagerange{\pageref{firstpage}--\pageref{lastpage}}

\maketitle

\begin{abstract}
In the star formation process, the vital impact of environmental factors such as feedback from massive stars and stellar density on the form of the initial mass function (IMF) at low-mass end is yet to be understood. Hence a systematic, highly sensitive observational analysis of a sample of regions under diverse environmental conditions is essential. We analyse the IMF of eight young clusters ($<$5 Myr), namely IC1848-West, IC1848-East, NGC 1893, NGC 2244, NGC 2362, NGC 6611, Stock 8 and Cygnus OB2, which are located at the Galactocentric distance ($R_g$) range  $\sim$6-12 kpc along with nearby cluster IC348 using deep near-IR photometry and Gaia DR2. These clusters are embedded in massive stellar environments of radiation strength $log(L_{FUV}/L_{\sun})$ $\sim$2.6 to 6.8, $log(L_{EUV})$ $\sim$42.2 to 50.85 photons/s, with stellar density in the range of $\sim$170 - 1220 stars/pc$^2$. After structural analysis and field decontamination we obtain an unbiased, uniformly sensitive sample of pre-main-sequence members of the clusters down to brown-dwarf regime. The lognormal fit to the IMF of nine clusters gives the mean characteristic mass ($m_c$) and $\sigma$ of 0.32$\pm$0.02 $M_\odot$ and 0.47$\pm$0.02, respectively. We compare the IMF with that of low- and high-mass clusters across the Milky Way.  We also check for any systematic variation with respect to the radiation field strength, stellar density as well with $R_g$. We conclude that there is no strong evidence for environmental effect in the underlying form of IMF of these clusters. 
\end{abstract}




\section{Introduction}
\label{introduction}

Molecular clouds provide the sites for stellar cluster formation. Their fragmentation results in the birth of stars over a broad spectrum ranging from high mass stars of several tens of solar masses to low-mass sub-stellar objects with masses below $0.07 M_{\odot}$. The mass of a star at birth is an important physical quantity as it determines the subsequent evolutionary path of the star and is a vital parameter in framing the star formation theories. The distribution of the stellar mass at birth, known as the stellar Initial Mass Function (IMF), is a fundamental property of star formation. In addition to enhancing our knowledge of the formation and evolution of stars, the shape of the IMF imposes a constraint on the star formation process (e.g. \citealt{krumholz2014,offner2014}) and it is a key quantity in many astrophysical studies. Therefore, a thorough understanding of its accurate shape is imperative (see various reviews by \citealt{bastian2010,jeffries2012,kroupa2013,offner2014,moraux2016}). 

The pioneering work of Salpeter in 1955 which formulated the power-law distribution for the high mass end of the IMF, led to the onset of a great deal of observational and theoretical studies, carried out to understand the form of IMF across different star forming environments. In recent years, the advancement of large telescopes have aided in extending these studies to the least massive stars and sub-stellar objects. The increasing focus in this field is substantiated by the various functional forms of the IMF available in the literature. For the Milky Way, \citet{salpeter1955} originally proposed a single power-law form for high mass side ($\ge$ 1$M_{\sun}$) of the IMF, which is generally approximated as $\frac{dN}{dM}$  $\propto$ $m^{-\alpha}$, with $\alpha$ = 2.35. More updated forms of IMF mainly include the multi-component power-law functions \citep{kroupa2001}, the tapered power-law form (\citealt{demarchi2005,demarchi2010}), and the log-normal distribution \citep{chabrier2003}. Although the above functional forms agree well with each other at the high mass end ($>$ 1 $M_{\sun}$), they deviate towards the low mass end (see \citealt{offner2014}) and as a result, we lack a model of star formation that predicts the IMF of a stellar population produced by a given molecular cloud.

Most of the observational studies on the IMF focus on the high-mass end which appears to be a fairly uniform distribution across the Milky Way disk and the local solar neighborhood (\citealt{bastian2010,offner2014,moraux2016} for review).  However, there have been  claims of non-universal IMF in the Milky Way as well as in external galaxies. \citet{Dib2014} through comparative analysis of mass function of eight galactic clusters using Bayesian statistics, states that the IMF is not universal. Similarly, non-universal IMF is observed in extreme environments such as in the galactic center (e.g. \citealt{lu2013,hosek2019}), in the most massive elliptical galaxies (e.g. \citealt{dokkum2010,cappellari2012}) and in the least luminous Milky Way satellites (e.g. \citealt{geha2013,gennaro2018}). \citet{guszejnov2019} showed that the current IMF models in literature either fail to reproduce the observed variations in the IMF of dwarf and elliptical galaxies or violate the universality of the IMF in the Milky Way. On the other hand, theoretical studies suggest that changes in the IMF depend on the local environmental conditions such as protostellar outflows, magnetic field, turbulence, or radiation feedback (see  \citealt{krumholz2016} and references therein). However, all these claims of a non-standard  IMF are highly debated still, which  demands for a more homogeneous approach to derive the  IMF for a sample of targets in diverse conditions. This uncertainty in the nature of the IMF and its potential dependence on the   environment poses to be one of the most challenging problems in modern astrophysics. Observations of the IMF of a variety of environments is essential to test the relative influence of various environmental factors such as gas temperature, stellar feedback and turbulence on the IMF (e.g. \citealt{hosek2019}).

In recent years, studies to explore the low-mass and sub-stellar end of the IMF have remarkably advanced, particularly in the nearby star forming regions (e.g. Taurus, IC348, 25 Ori; see \citealt{luhman2016,luhman2018,suarez2019}). Unfortunately, these nearby ($<$ 500 pc) environments are poor analogues of the diverse star forming conditions, where metallicity varies or massive stellar feedback dominates. Moreover, most of the nearby star forming regions contain mainly low-mass stars with no or very few massive stars -- with the exception of the ONC, which is the only massive star forming region within 500 pc. On the other hand, most of the regions with embedded massive stars are located further away and their IMF determination extends usually down to a few solar masses only, and in a few selected cases reaches mass close to the hydrogen burning limit. For e.g., the studies on Westerlund 1 by \citet{andersen2017} and on Trumpler 14 by \citet{rochau2011} extended down to $\sim0.15 M_{\sun}$, \citet{muzic2017} estimated the IMF in RCW38 down to 0.02 $M_{\sun}$ while the IMF has been characterized down to 0.4 $M_{\sun}$ in NGC3603 (\citealt{stolte2006,harayama2008}). However, there still lack a systematic, uniform analysis of low-mass stellar IMF of a statistically rich sample of Galactic young clusters of diverse properties and  located at various environmental conditions.  

The study of low-mass IMF is extremely challenging due to the lack of complete and clean sample of young members which is an absolute essential for its accurate estimation.  Also, studies on individual regions are often biased by different sources of uncertainties such as the use of different evolutionary models, reddening laws, membership criteria, non-uniform sensitivity of observations rather than the variation between individual clusters \citep{muzic2019}. With the understanding that stars in young clusters have roughly the same age, metallicity, and are located at the same distance, the presumption that their observed present day mass functions (PDMFs) are a fair representation of their IMFs seems reasonable. In particular, rich or moderately rich young cluster sample is of particular importance for IMF studies since such clusters host thousand to a few thousand stars, and are better for robust statistical measurements. Moreover such young clusters of a few Myr age are not expected to be dynamically evolved. The effect of mass segregation due to dynamical evolution is more pronounced in old clusters and can preferentially remove low-mass members from cluster center. Young clusters are largely free of the effects of dynamical evolution although primordial mass segregation is expected up to some extent. 

In order to explore the role of environmental factors in the form of the low-mass stellar IMF, in this study, we have selected a sample of young clusters of diverse properties in terms of their  stellar density, number of associated massive stars and located at various  Galactocentric distance. The environmental conditions in these regions are significantly different from those in nearby star forming regions. The main focus of this study is to explore the low-mass part of the IMF (i.e., $<$ 3 $M_{\sun}$ and  down to brown-dwarf regime) by obtaining an unbiased, uniformly sensitive sample of members in these regions and to correlate it with the well studied mass distributions in nearby star forming regions and young massive clusters. The selected clusters are young enough for dynamical processes and stellar evolution not to have significantly altered the stellar mass distribution and the physical conditions of the birth environment have not been completely erased. In this work, we do not perform corrections for binary stars, as it might not affect the overall shape of the IMF significantly. The studies in young clusters  by \citealt{harayama2008,zeidler2017,muzic2017,suarez2019} show that the  effect of binarity in IMF calculation is not very pronounced as single-star IMF and the system IMF agree within errors. 

The  paper is structured as follows. In Section 2, we describe the various properties of the sample clusters in our study and the NIR photometry data sets used for the analysis. Section 3 explains the estimation of cluster fundamental parameters such as center, radius, distance, reddening and age, field star decontamination process and membership criteria leading to the estimation of IMF. Section 4 discusses the IMFs estimated for the regions under study and their comparison with other well-studied regions. Section 5 summarizes the various results obtained.

\section{Sample Selection and Data}

\subsection{Sample Selection}
\label{sample}

\begin{figure*}
 \centering
 \includegraphics[scale = 0.25]{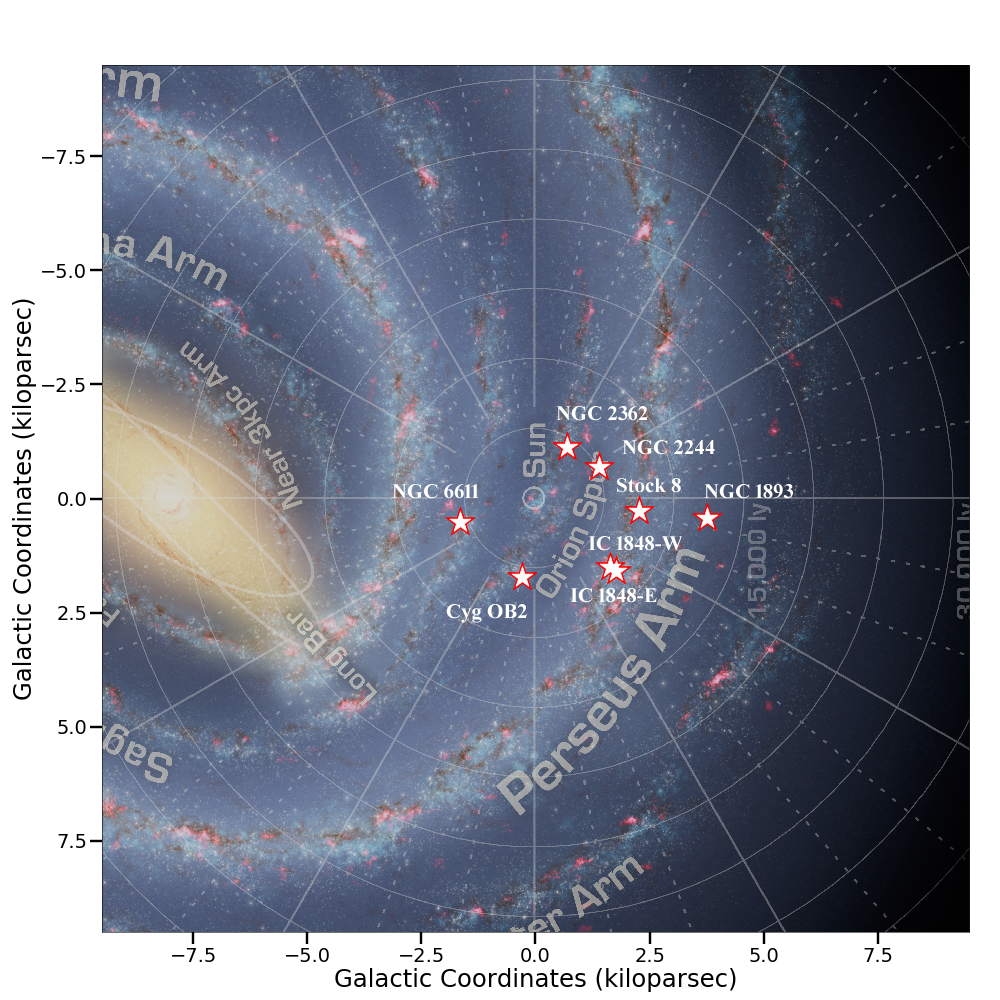}
 \caption{Spatial distribution of the clusters in the study  with respect to the Sun and the Galactic centre. $R_o$ (distance of the Sun from the Galactic centre) is taken to be 8.34 kpc \citep{reid2014}. The image is adapted from https://www.universetoday.com/102616/our-place-in-the-galactic-neighborhood-just-got-an-upgrade/ , credit by Robert Hurt, IPAC; Bill Saxton, NRAO/AUI/NSF.}
 \label{fig:galactic_plane}
\end{figure*}

To make a comprehensive statistical analysis of the form of IMF in diverse environmental conditions, we have selected eight young  clusters from various studies (\citealt{chauhan2011,king2013,jessy2017,kuhn2019}). 
We select the clusters satisfying the following criteria i) age $<$ 5 Myr with massive O,B stars at its center ii) rich population of PMS stars  with relatively low interstellar reddening iii) availability of deep JHK data from 4m class telescopes iv) spatially distributed over a wide range of Galactocentric distance.  The clusters included in this work are IC1848-East, IC1848-West, NGC 1893, NGC 2244, NGC 2362, NGC 6611, Stock 8 and the bright cluster at the center of Cygnus OB2 association (which we refer as  Cygnus OB2 hereafter). The details are given in Table~\ref{tab:cluster_list}. Below we discuss the individual clusters, the massive stars associated with them, their basic characteristics such as age, distance, reddening etc.  Fig.~\ref{fig:galactic_plane} shows the spatial distribution of the clusters in this study with respect to the Sun and Galactic center, where $R_o$, the Galactocentric distance of the Sun is taken to be 8.34 kpc \citep{reid2014}. The JHK colour composite images of the clusters are given in Appendix~\ref{sec:FOV_of_clusters}.

\subsubsection{IC1848-West and IC1848-East}

Westerhout 5 or W5 is one of the three major star forming clouds in the W3/W4/W5 giant molecular cloud (GMC) of Cassiopeia OB6 association (see \citealt{jessy2016}). W5 has two prominent \hii regions associated with the clusters IC1848-West and IC1848-East. Both clusters have massive stars at their centre ionising the environment around them by strong UV radiation \citep{koenig2008}. Hence these clusters serve as ideal targets to study the star formation process within the  feedback environment of  massive stars.

The cluster IC1848-West has two O stellar  groups dominating at its center (HD17505 and HD17520) and each one is a multiple system. The multiple system HD17505 contains at least four O stars (O6.5III((f)), O7.5V((f)), O7.5V((f)), O8.5V) 
that are apparently gravitationally bound (\citealt{hillwig2006,raucq2018} and references therein). The binary system HD17520 is associated with an O8V and O9:Ve stars \citep{sota2011}. The cluster IC1848-West is located at a distance of 2.2 $\pm$ 0.2 kpc \citep{moffat1972,1848lim2014} with the mean interstellar reddening $A_v$ =  2.05 $\pm$ 0.17 mag \citep{1848lim2014}. Using  PMS evolutionary models for  low-mass stars, an average age  of $\sim$ 3 Myr have been estimated in this cluster \citep{1848lim2014}.    

IC1848-East is primarily ionised by HD18326, a binary system of spectral types O6.5V((f))z and O9/B0V \citep{sota2014} located at the centre of the cluster. The $A_V$ in the cluster is in the range of $\sim$ $1.9-2.5$ mag with a mean age of young stellar objects as $\sim$ 1-2 Myr and is located at a distance of $\sim$ 2.1 $\pm$ 0.3 kpc \citep{chauhan2011}.

\subsubsection{NGC1893}

NGC1893 is a young open cluster embedded in the IC410 \hii region at the centre of the Auriga OB2 association located towards the Galactic anti-centre. There are five O type stars (HD 242926 - O7V; HD 242908 - O4V((f)); LS V +34\textdegree15 - O5.5V((f)); BD +33\textdegree1025A - O7V; HD 242935 - O7.5V((f))) embedded in the cluster \citep{negueruela2007}. Previous studies estimate the cluster to be at a distance of $\sim$ $3-6$ kpc (\citealt{tapia1991,marco2001,sharma2007,prisinzano2011,pandey2013}). The mean  $A_v$ of the cluster is in the range of $\sim$ 1.5-1.9 mag (\citealt{sharma2007,prisinzano2011,lim2014}) and is $\sim$ 1.4-1.9  Myr old (\citealt{prisinzano2011,lim2014}). Though located at a far distance, the population of massive stars and relatively low interstellar reddening make this region an ideal target to study the effect of external factors on star formation \citep{negueruela2007}.

\subsubsection{NGC2244}

NGC 2244 is a young cluster associated with the star forming region Monoceros OB2 (Mon OB2) association. This system is located in the northwest quadrant of the Rosette Molecular Cloud complex, which is one of the most massive GMCs in the Milky Way \citep{chen2007}. The cluster houses seven massive O type stars (HD46223 - O4V((f)); HD46150 - O5V((f))z; HD46485 - O7Vn; HD46056 - O8Vn; HD46149-1 - O8V; HD46149-2 - O8.5-9V; HD46202 - O9.5V) along with numerous B type stars which ionise the neighbourhood \citep{martins2012}. The estimated distance to the cluster is in the range of $\sim$  $1.3-1.7$ kpc (\citealt{perez1987,hensberge2000,park2002,muzic2019}). The  $A_v$  is $\sim$ $1.4-1.7$ mag (\citealt{massey1995,li2005,ngc2244bonatto2009}) and   average age is $\sim$ $2-3$ Myr (\citealt{perez1991,hensberge2000,ngc2244bonatto2009}) for the cluster.

\subsubsection{NGC2362}

NGC 2362 is a young star cluster located in the third Galactic quadrant, dominated by the 4th mag O9 Ib multiple star $\tau$ CMa and nearly three dozen B-type stars distributed in a volume of $\sim$3 pc  radius \citep{dahm2007}. Using optical photometry,  \citep{moitinho2001} obtained the distance to the cluster as $\sim$ 1.5 kpc and an age of $\sim$ $3-6$ Myr. They also show that the cluster has a well defined locus of PMS stars in the colour-magnitude diagram which makes it an ideal laboratory for stellar evolution studies. Considering the young age of the cluster, the region has a rather low, uniform  $A_v$ of  $\sim$ 0.31 mag (\citealt{moitinho2001,dahm2005,delgado2006}).

\subsubsection{NGC6611}

The young cluster NGC 6611 is part of the Ser OBI association in the Sagittarius spiral arm and is responsible for ionizing the well known nebula M16 (The Eagle Nebula) in the W37 molecular cloud \citep{hillenbrand1993}. This is the only cluster directed towards the Galactic center in our list. Structurally the region has several  nebular features, the so-called elephant trunks (Pillars of Creation) and  at the tips of which new-born stars are visible \citep{ngc6611bonatto2006}. There are 13 O type stars embedded in the cluster (\citealt{hillenbrand1993,evans2005}). The distance to the cluster is estimated to be in the range of $\sim$ $1-2$ kpc (\citealt{hillenbrand1993,dufton2006, ngc6611bonatto2006,ngc6611guarcello2007}). Various studies have shown that the cluster is of $\sim$  $1-4$ Myr old (\citealt{hillenbrand1993,dufton2006, ngc6611bonatto2006}) and the  extinction in the region varies in the range of $\sim$ 1.4 - 3.1 mag \citep{oliveira2009} with an average value of $\sim$ 2.6 mag \citep{ngc6611guarcello2007}. 

\subsubsection{Stock8}

Stock 8 is located within the \hii region of IC 417 (Sh2-234) in the Auriga constellation of the Perseus arm. It is surrounded by 12 massive OB type stars and is probably part of a large OB association (\citealt{jessy2008,marco2016,jessy2017}). Various distance estimation of the cluster lies in the range  of $\sim$  2.05 to 2.8   kpc  \citep{jessy2008,marco2016}. The reddening within the cluster region ($A_v$) has been estimated  in the range of  $\sim$ $1.2-1.9$ mag and age between 1 and 5 Myr \citep{jessy2008}. Using PMS evolutionary model fitting, \citet{jessy2017} obtains a median age of $\sim3.0$ Myr with an  age spread of $\sim$ 2 Myr for the cluster. The star LS V +34$^\circ$23 with spectral type O8 II(f) is likely to be the main source of ionization of the \hii region \citep{marco2016}.

\subsubsection{Young cluster within Cygnus OB2}

Cygnus OB2 (Cyg OB2) is a young massive OB association in the Cygnus X region \citep{winter2019}. It contains many massive stars up to $\sim100 M_{\sun}$ \citep{wright2015} which contribute to the strong FUV radiation fields in the region. Based on infrared studies, \citep{knodlseder2000} estimated the total number of O type stars within Cygnus OB2 as $120\pm20$, claiming to be the largest population of O stars found in a Galactic massive stellar association. The $A_v$ of  member stars range from $\sim$ $5-20$mag \citep{knodlseder2000}. Based on the most massive dwarf stars in the bulk of the cluster, \citet{hanson2003} estimates an average age of $\sim$ 2 Myr with a spread of 1 Myr. However, \citet{wright2015} suggest that  majority of star formation in Cyg OB2 occurred more or less continuously between 1 and 7 Myrs ago.  Using the Gaia parallax and fitting a 2-component Gaussian model, \citet{berlanas2019} obtained the median distance to the cluster as 1755 pc. \citet{bica2003} identifies two young clusters towards the center of the rich compact association Cygnus OB2, named as Object 1 and 2, and in this study, we analyse the southern cluster Object 1.

\subsection{Deep NIR photometry}
\label{subsec:deep_nir_photometry}

We have gathered deep NIR photometry in J, H and K-bands from various sky surveys and observations taken using 4 m class telescopes. Our main goal is to obtain uniformly sensitive deep photometry data sets for all the regions in order to sample the stellar masses down to $\sim$ 0.08 M$_{\sun}$. Field of view of all the  regions in JHK bands are given in Appendix~\ref{sec:FOV_of_clusters}. Below we describe the various data sets used for individual clusters and are also listed in Table~\ref{tab:cluster_list}.

For the clusters NGC2244, Stock8 and Cygnus OB2, we took the photometry from UKIDSS (United Kingdom Infrared Deep Sky Survey; \citealt{ukidss_vosa}) DR6 GPS (Galactic Plane Survey) observed using the WFCAM (Wide Field Camera) on the 3.8m UKIRT (United Kingdom Infrared Telescope). We limit the catalogue to sources with goodness of fit (pstar) $>$ 0.9 which gives the probability of the source being a star and PriOrSec(m) = 1 which removes the duplicated sources located in the overlapping regions between different arrays in a WFCAM tile \citep{lucas2008}. 

For the clusters NGC1893, NGC2362 and NGC6611, we use the photometry from the MYStIX survey (Massive Young Star-Forming Complex Study in Infrared and X-ray; \citealt{feigelson2013}) observed using the WFCAM on the UKIRT. We restrict our catalogue to sources with J, H and K magnitudes flagged as 'O'. This constraint removes the bad-pixels  and non-stellar sources \citep{king2013}.   Apart from the individual source selection criterion for the various catalogues, to ensure photometric accuracy, we use only those sources with photometric uncertainty within 0.2 mag in J, H and K-bands. 

For the clusters in W5 complex (IC1848-East and IC1848-West), we obtain the data from the NOAO archive\footnote{\url{http://archive1.dm.noao.edu/search/query/}}. The observations in J, H and K-bands were conducted using the wide-field IR imager NEWFIRM (NOAO Extremely Wide Field Infrared Imager; \citealt{probst2004}) with the 4m Mayall Telescope at Kitt Peak National Observatory, Arizona (PI: Guy Stringfellow). The NEWFIRM camera includes four InSb 2048 $\times$ 2048 number of pixel arrays arranged in
a 2$\times$2 pattern and the  field of view is 28$^\prime$ $\times$ 28$^\prime$ with a pixel scale of 0.4$^{\prime\prime}$. Using the NEWFIRM Science Pipeline \citep{swaters2009},  the standard processing of dark correction, flat fielding, sky subtraction and bad pixel masking were performed and we obtained the final calibrated, stacked and mosaiced images in three bands from the archive. The FWHM of the images were in the range of 0.8$^{\prime\prime}$ - 1.0$^{\prime\prime}$. Using DAOFIND task in IRAF we obtained the list of point sources in $K$-band with signal 5$\sigma$ above the background. The 5$\sigma$ detection criteria was useful to avoid any false detection or artefacts in the image. The list of sources were again visually checked to exclude any spurious detection. The same source list was used for $J$ and $H$-bands as well. We performed psf photometry of these sources using the ALLSTAR routine of IRAF (eg. \citealt{jessy2016,jessy2017}).  For absolute photometric calibration, we used the Two Micron All Sky Survey (2MASS) catalog \citep{cutri2003} of those sources with quality flag  `A' in all the three bands. The match radius used to obtain the common sources in the 2MASS and NEWFIRM catalogs was 1.0$^{\prime\prime}$. The zero point correction term with respect to 2MASS photometry have been applied to our NEWFIREM photometry for individual bands in order to calibrate it. The calibration accuracy was within 0.05 - 0.07 mag for all three bands. Our final NEWFIRM photomerty list includes only those sources with  S/N$>$5 and photometry uncertainty $<$ 0.2 mag in all three bands.

For the above clusters, we have examined $\sim$ $20\arcmin \times 20\arcmin$ area and a nearby control field (see section \ref{sec:fsd} for details)  to study the cluster properties. In general, our photometry has a wide dynamic range of $\sim$ 12--20 mag in J-band. The saturation limits of individual data set differ depending on various factors such as sky background, seeing, exposure time etc. across different observations. Considering the saturation limit of UKIDSS photometry, we exclude sources brighter than 13 mag in J band \citep{lucas2008} for our analysis. Since this study focuses only on the low-mass end of IMF ($<$ 3 $M_{\sun}$), this cut off will not affect our results.

\begin{table*}
	\caption{Details of the sample clusters }
	\label{tab:cluster_list}
	\begin{tabular}{llcccccccccc}
		\hline
		\multirow{3}{*}[-3pt]{Sample} & \multirow{3}{*}[-3pt]{Data Set} & \multicolumn{2}{c}{Cluster} & \multicolumn{2}{c}{Control Field} & \multicolumn{6}{c}{Data completeness}\\
		\cmidrule{3-12}
		& & RA & Dec & RA & Dec & J band &  H band & K band & J band & H band & K band\\
		& & (deg) & (deg) & (deg) & (deg) & (mag) & (mag) & (mag) & (M$_{\sun}$) & (M$_{\sun}$) & (M$_{\sun}$)\\
		\hline
		IC1848-West & NEWFIRM & 42.7958 & +60.4019 & 44.9642 & +60.7584 & 19.0 & 18.5 & 18.5 & 0.04 & 0.04 & 0.03\\
		& (Mayall Telescope) &\\
		IC1848-East & NEWFIRM & 44.8458 & +60.5667 & 44.9642 & +60.7584 & 18.5 & 17.5 & 17.0 & 0.08 & 0.11 & 0.11\\
		& (Mayall Telescope) &\\
		NGC1893 & WFCAM & 80.7064 & +33.4273 & 80.6741 & +33.5531 & 18.5 & 17.5 & 17.5 & 0.19 & 0.22 & 0.17\\
		& (UKIRT, MYStIX)&\\
		NGC2244 & WFCAM & 97.9808 & +04.9431 & 97.9842 & +05.1100 & 17.5 & 17.0 & 16.5 & 0.07 & 0.07 & 0.07\\
		& (UKIRT, GPS) &\\
		NGC2362 & WFCAM & 109.6865 & -24.9582 & 109.3938 & -24.6989 & 19.0 & 18.5 & 18.5 & 0.02 & 0.02 & 0.02\\
		& (UKIRT, MYStIX) &\\
		NGC6611 & WFCAM & 274.6700 & -13.7900 & 274.4976 & -13.7934 & 19.0 & 18.0 & 17.0 & 0.04 & 0.05 & 0.07\\
		& (UKIRT, MYStIX) &\\
		Stock8 & WFCAM & 82.0373 & +34.4244 & 82.2884 & +34.5069 & 18.5 & 18.0 & 17.5 & 0.09 & 0.08 & 0.09\\
		& (UKIRT, GPS) &\\
		Cygnus OB2 & WFCAM & 308.2827 & +41.2155 & 308.5783 & +41.4213 & 18.5 & 17.5 & 17.0 & 0.12 & 0.10 & 0.09\\
		& (UKIRT, GPS) &\\
		\hline
	\end{tabular}
\end{table*}

\section{Analysis and Results}

Section \ref{sample} shows that numerous studies exist in the past to examine the various physical parameters such as radius, distance, reddening and age of each cluster. Most of those analysis show a large range in the parameters estimated. However, a uniform method to analyse these properties is essential in order to exclude any bias in the estimation of IMF. In this section we use the deep JHK photometry to analyse the various physical properties of the clusters.

\subsection{Data completeness}

\begin{figure*}
    \centering
    \includegraphics[width=\textwidth]{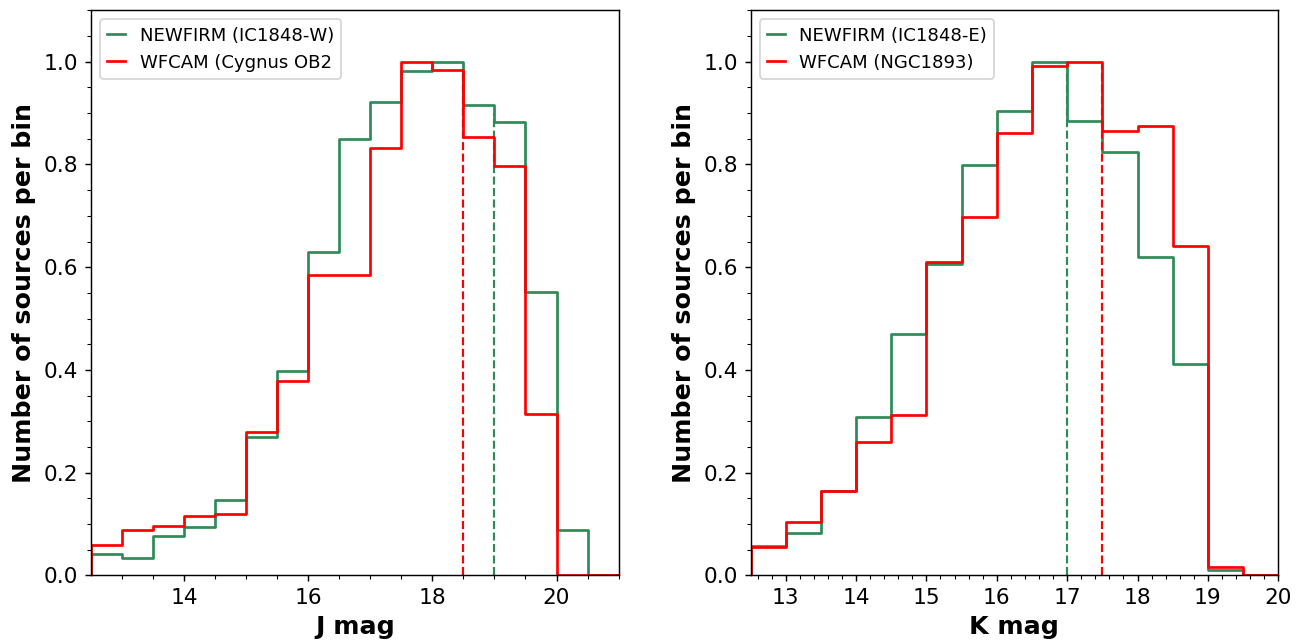}
    \caption{Normalized sample histograms showing the completeness limits of the data used from different telescopes in J band (left panel) and K band (right panel).  The turnover point in the distribution serves as a proxy for the completeness limit of the data. The dashed lines mark the $\sim$ $90\%$ completeness limit of the photometry.}
    \label{fig:completeness_hist}
\end{figure*}

Photometric data often suffer from data incompleteness mostly towards the fainter end. There are various factors contributing to this such as sensitivity of different observations, crowding and variable extinction. In order to evaluate the completeness of the photometry used in this analysis, we plot histograms of the sources detected within the area considered for the analysis for each cluster.  The turnover point in the histogram distribution is generally considered as the pointer for $\sim$ 90 \% completeness (e.g. \citealt{willis2013,samal2015,maia2016,jessy2017}). In the cases where the bin trailing the turnover point is more than 90 \% of the peak value, then that magnitude is taken as the completeness limit \citep{jessy2016}. The $\sim$ 90 \% completeness limits thus estimated for all the clusters in this study are given in Table~\ref{tab:cluster_list}. Fig.~\ref{fig:completeness_hist} shows the sample completeness histograms in J (left panel) and K-bands (right panel) of two different clusters, one for each of the telescopes used in this study (i.e. UKIRT and Mayall telescopes) (see section~\ref{subsec:deep_nir_photometry}). The completeness histograms corresponding to the control fields of these clusters are represented in Fig.~\ref{fig:cf_completeness_hist} (see appendix~\ref{sec:cf_compl_hist}). From Figs.~\ref{fig:completeness_hist} and ~\ref{fig:cf_completeness_hist} it is clear that the completeness of the cluster and control fields are comparable. Our estimates for the completeness of the data in all the three bands agree with the 90 \% completeness limits mentioned in the UKIDSS GPS catalog details \citep{lucas2008}. 

The above method of stellar counting for data completeness measurements should be appropriate for most of the area in the cluster. However, the local stellar surface density and/or the presence of bright stars within the cluster region can have an effect on the completeness of the data. (eg. \citealt{maia2016,andersen2017}). A more rigorous analysis of data incompleteness by the artificial star simulations by \citet{maia2016} and \citet{jessy2017} show that the data incompleteness from stellar count method and artificial star simulations are correlated for most of the area of interest. However, photometry can be incomplete  towards the cluster center in extreme crowded regions (e.g. \citealt{maia2016}). In order to analyze any spatial variation, we measured the completeness by the artificial star method within annular radii of 1 arcmin from the center of Stock 8 to the  outward. The completeness varies from $\sim$ 85 to 95 \% from inner to outer radius of the cluster with an average of $\sim$ 90 \% for the magnitude bin 17-18 in K-band, which is in agreement with our histogram method (see Table~\ref{tab:cluster_list}). In this paper we do not account for the spatial variation of incompleteness across the cluster regions. We use the photometry down to the 90 \% completeness limits given in Table~\ref{tab:cluster_list} for follow-up analysis. The corresponding mass completeness limits estimated using \citet{baraffe2015}  evolutionary models after  incorporating the distance, reddening and age of the respective clusters  are also given in Table~\ref{tab:cluster_list} (see sections~\ref{sec:dist_est},\ref{sec:ext},\ref{sec:age} for details). In general, our photometry is complete down to $\sim$ 0.08 $M_{\sun}$ for most of  the regions except NGC1893.

\subsection{Cluster center and radius}
\label{sec:center&radius}

One of the critical steps in membership analysis and IMF construction is determining the radial extent of the cluster. To delineate the cluster area we determine the stellar surface density and consider the central region with density higher than that of the surroundings, (which includes both, the foreground and background field stars) as the cluster limit. There are two basic approaches for density estimation, parametric and non-parametric techniques. If the cluster probability density function is known then parametric methods are useful. Since this is unlikely, we rely on non-parametric methods \citep{srirag2019}. We employ two non-parametric approaches to determine the cluster radius: star count method and k-nearest neighbours method.

\subsubsection{Star count method}

One of the commonly used methods to determine the cluster radius is by analysing the variations in the stellar surface density (eg. \citealt{baba2004,ojha2011}). In this regard, the star count method is a simple yet robust approach. In this method, we grid the sample region into bins of equal size and estimate the individual bin density, i.e. the number of sources in each bin divided by the bin area.  For uniformity, we consider a bin size of 0.3 pc for all the clusters under study (i.e, the angular sizes of the bins varied from 15 to 45 arcsec depending on the distance to the cluster). Choosing the appropriate bin size is critical since a small bin prohibits a meaningful statistics and a large bin hides the underlying cluster features. After various trials using different bin sizes, 0.3 pc seems to be ideal for both the nearby and distant clusters. Fig.~\ref{fig:star_count_nn_method} shows a Hess diagram of the spatial density distribution of a sample cluster, IC1848-West. The density in each bin is smoothed over with nearby bins by interpolation and the  colour gradient indicates the density variation across the cluster area. In order to obtain the background density, we consider a control field for each cluster which is at the same Galactic latitude as that of the cluster. The coordinates of the control field for each cluster are given in Table~\ref{tab:cluster_list}. We measure the mean ($\mu$) and standard deviation ($\sigma_{bg}$) values of the stellar density distribution within the control field.   The cluster is assumed to lie within a region defined by  $\mu$ + $3 \sigma_{bg}$. We restrict the radius to 3$\sigma$ in order to reduce the field contaminants as well as to avoid the differential reddening in a larger area. The radius of the circle covering this area and the highest density bin within it are taken as the cluster radius and centre, respectively. The cluster radius obtained through this method is tabulated in Table ~\ref{tab:phy_para}.

\subsubsection{Nearest Neighbours method}

\begin{figure*}
    \centering
    \includegraphics[scale = 0.75]{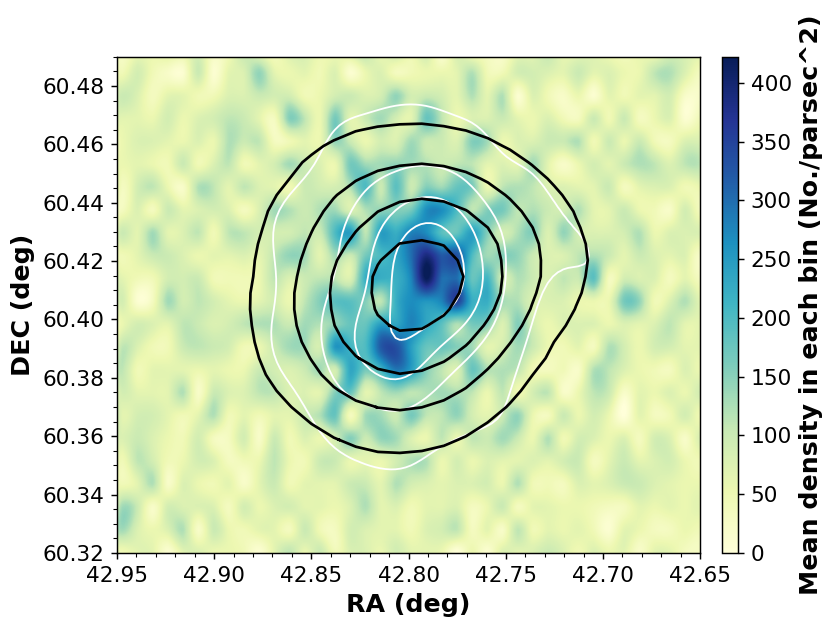} 
    \caption{Surface density plot of sample cluster IC1848-West. Black contours and white contours mark the $3\sigma$, $5\sigma$, $7\sigma$ and $9\sigma$ levels above the background density estimated by the nearest neighbours method and star count method respectively.}
    \label{fig:star_count_nn_method}
\end{figure*}

Unlike the star count method, which depends on the density gradient between the cluster and the background field, the k- nearest neighbours method is a more reliable and well-acclaimed method to estimate the stellar surface density. The method introduced by \citep{casertano1985} gives the generalized form of the $j^{th}$ nearest neighbour surface density for a star as:
\begin{equation}
 \rho _j = \frac{j-1}{\pi r_j^2}
\end{equation}
where $r_j$ is the distance from any given star to its $j^{th}$ neighbour. Similar to star count method where choosing the appropriate bin size is critical, in this method, using the right j value is essential. If the j value is too small then insignificant sub-clustering or false groupings become prominent. At the same time if j is large then small scale high density subgroups will be overlooked. Hence for detecting substructures within a cluster a lower j value is preferable, while higher j values may be used to trace large-scale structures \citep{schmeja2011}. For our analysis, we varied the values of j as 10, 15 and 20, and j=15 was found to be optimal to trace the radius of the clusters. Similar to the star count method, we estimate the background counts ($\mu$ and $\sigma$) by averaging the density of the control field. In Fig.~\ref{fig:star_count_nn_method} we overplot contours for the $3\sigma$, $5\sigma$, $7\sigma$ and $9\sigma$ levels above the background density estimated by this method. The radius of the $3\sigma$ contour has been considered as the cluster radius and the location of highest density as the cluster center. The cluster radius obtained through this method is tabulated in Table ~\ref{tab:phy_para}.

The radius estimated using the star count method is consistent with that estimated using the nearest neighbours method for all the clusters except NGC1893. NGC1893 is an elongated star forming region, with two prominent cometary globules towards its north-east, making the outer edge of the cluster significantly reddened \citep{sharma2007}. In order to avoid these features we adopt the cluster radius as  $3\arcmin$ from \citet{sharma2007}. The mean of the cluster center and radius measured from star count and nearest neighbours method have been adopted as the cluster center and radius for each cluster. The adopted central coordinates and the radius of each cluster are listed in Tables~\ref{tab:cluster_list} and ~\ref{tab:phy_para}, respectively. For further analysis, we use the data within the radius of each cluster. In  young clusters  majority of their stars lie within a few core radius (see \citealt{sharma2007,jessy2008}), thus effect of halo stars lying beyond the radius adopted here on the IMF estimation is unlikely to be significant. We discuss this point in section~\ref{imf}.

\subsection{Field star decontamination}
\label{sec:fsd}

Field stars in astronomical studies are those which do not belong to the celestial system being observed. They can be the foreground or background sources which appear in the same field of view as the object. An initial step in our IMF analysis is to identify the probable cluster members by removing the field star contaminants from the data sets. This membership determination is crucial for estimating the IMF because both PMS and dwarf foreground stars overlap in the colour-magnitude diagrams (CMDs; e.g. \citealt{jessy2017}). There are various methods for assessing the membership probabilities of the sources in a cluster utilizing the positions, proper motions, radial velocities, spectroscopy or multi-band photometry of the stars (e.g. \citealt{panwar2017,datta2018,herczeg2019,bharadwaj2019,jose2020} and references therein). It is commonly accepted that precise membership probabilities are obtained by using kinematic parameters of the sources. However, our clusters lie beyond the solar neighbourhood and astrometric data in general tend to have high uncertainties towards fainter end and hence constrain our options of low-mass membership analysis. Without kinematic or spectroscopic data for the clusters, a main method to separate the members from the field stars is by statistical subtraction using a nearby control field (e.g. \citealt{jessy2017,kaur2020}). For all the clusters in this study, we choose a control field of similar area in the cluster vicinity (see Table~\ref{tab:cluster_list}). The control field for each cluster was selected based on the assumption that it is not located far away from the cluster, so that its Galactic field stellar distribution would be similar as that in the cluster. However, it cannot be too  close to the cluster center, which may include some of the cluster members. We avoided regions with too much nebulosity which is a signature of active star formation. We also excluded the regions with young stellar objects associated with these clusters which are listed in the previous studies (Section~\ref{sample}). The control fields thus selected are $\sim$ 7 to 22 arcmin away from the cluster  center (see  Table~\ref{tab:cluster_list}). As the clusters IC1848-East and IC1848-West lie in the same star forming complex, we consider the same control field for both of them and is taken from the corner of the NEWFIRM image of IC1848-East. Since the control fields are relatively close to the cluster regions, the reddening variation between the two regions are negligible and hence we assume that similar amount of field contaminants exist in the cluster region as well as in the control field. In order to validate this assumption of similar extinction distribution both for the cluster and control fields, in the panel 3 of figures given in Appendix~\ref{sec:fd_CMD_of_all_clusters}, we have over plotted the CMDs  of cluster and control field regions. It is evident from all the figures that the control field sequence exactly overlays on the field sequence of the cluster region and there is no significant offset between the two distributions. This shows that the extinction differences between the cluster and control fields are negligible and hence we could perform statistical subtraction without giving any additional correction in the extinction values of the control field.

\begin{figure*}
    \centering
    \includegraphics[width=\textwidth]{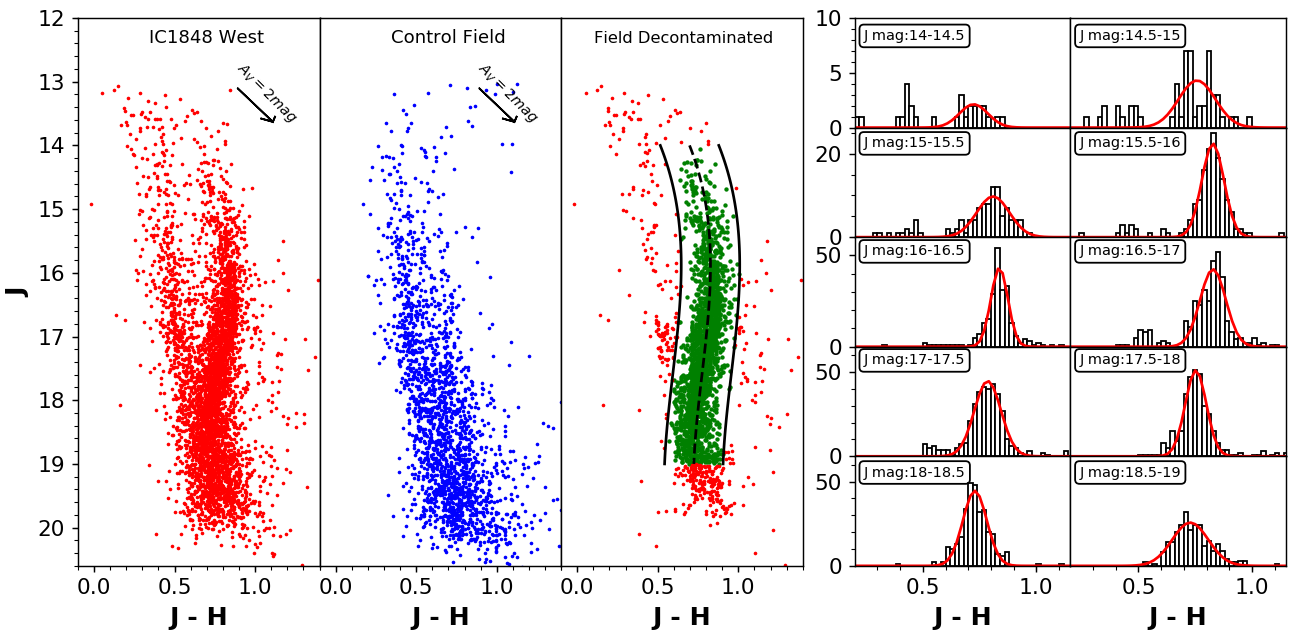}
    \caption{{\it Left}: J-H vs J CMD of sample cluster IC1848-West for all the sources within the radius (left panel), nearby control field (middle panel), CMD after field decontamination (right panel). Reddening vector for Av=2mag is also shown. {\it Right}: Histograms showing the distribution along J-H axis for every 0.5 mag strip of J band. The mean of the Gaussian fit (red curve) gives the mean locus of the PMS population in that magnitude bin. The dashed curve in field decontaminated CMD (left) represents the interpolation of PMS locus of each bin and the continuous curves mark the $3\sigma$ deviation from the mean locus. The candidate PMS members within $3\sigma$ distribution are highlighted in green.}
    \label{fig:fd_cmd}
\end{figure*}

The field stars were statistically removed by the following steps. We use the (J-H) vs J CMD for both the cluster as well as the control field regions  within the radius determined from the section~\ref{sec:center&radius}. We prefer this combination because J and H bands are least affected by NIR excess emission from circumstellar disk around young stellar objects. We then divide the colour and magnitude axes into bins of size 0.1 and 0.2 mag, respectively. For every source in each bin of the control field, the corresponding source in the cluster region is considered as a background star and is removed. By repeating this process for all the bins in the field CMD, we obtain the background subtracted sources in the target cluster. In Fig.~\ref{fig:fd_cmd}, we  show the (J-H) vs J CMDs of a sample cluster IC 1848-West and its control field. The CMDs of all other clusters are given in Appendix~\ref{sec:fd_CMD_of_all_clusters}. A comparison of cluster and field region CMDs in Fig.~\ref{fig:fd_cmd} shows that the sequence of stars seen on the left side in both figures is the field stellar population and the additional sequence towards the right in the cluster CMD is the locus of the candidate PMS members of the cluster (see \citealt{jessy2017}). The PMS branch is well defined and lies separated from the field star distribution between $\sim$14-18 mag in J-band and merges with the field sequence at fainter magnitudes, as the low mass PMS sources have an intrinsic blueish colour as evident in the PMS models \citep{baraffe2015}. All the clusters in our list have similar well defined PMS branch (see Appendix~\ref{sec:fd_CMD_of_all_clusters}). Also, the  PMS branch has relatively narrow  distribution of points, revealing that the differential extinction within the clusters are minimal (see section~\ref{sec:ext} for details), unlike other very young clusters that are still embedded in molecular clouds (e.g., \citealt{jessy2016,panwar2017,muzic2017,bik2019}). The width of the dispersion of points in colour may be due to variability, binarity, age spreads or  photometric uncertainties (see \citealt{jessy2017,kuhn2017} for details). In the right panel of Fig.~\ref{fig:fd_cmd}, the field decontaminated (i.e.,statistically subtracted) CMD of the cluster IC1848-West is shown (see Appendix~\ref{sec:fd_CMD_of_all_clusters} for rest of the regions). 

\begin{figure*}
    \centering
    \includegraphics[width=\textwidth]{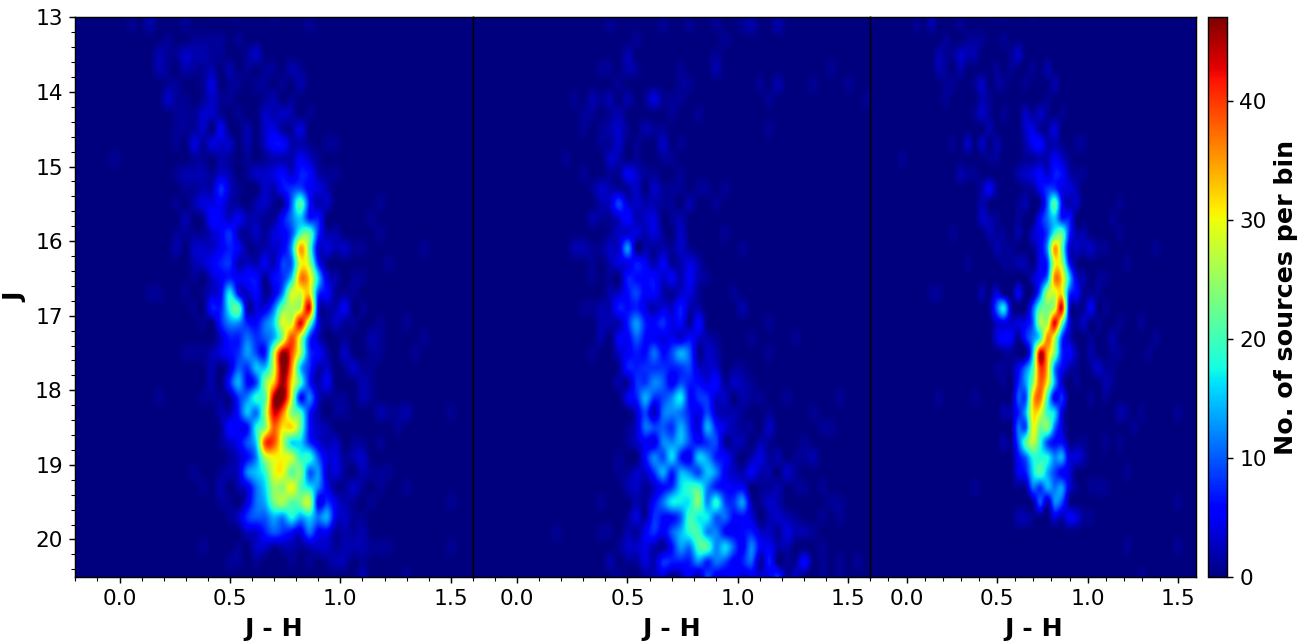}
    \caption{Hess diagrams of J-H vs J CMD of sample cluster IC1848-West (left), control field (middle) and cluster region after field star decontamination (right). The colour bar indicates the number of sources in each bin of size 0.04 along (J-H) and 0.2 along J.}
    \label{fig:sample_sel_cmd}
\end{figure*}

In Fig.~\ref{fig:sample_sel_cmd}, hess diagrams (which shows the relative density distribution of stars in the CMD) of the cluster IC1848-West (left panel), control field (middle panel) and the field star decontaminated cluster region (right panel) are shown. The Hess diagrams clearly shows a density enhancement along the PMS branch that remains the same after field decontamination. 

The right panel in Fig.~\ref{fig:fd_cmd} shows the distinct presence of the PMS population along with a few scattered sources. These scattered sources are most likely field contaminants which were not removed due to the statistical uncertainty in our decontamination process. In order to refine the decontamination process and clean our final catalogue, we define the locus of the PMS branch by fitting a Gaussian function along the colour axis and for bins of 0.5 magnitude in J band.  Since the PMS branch is roughly vertical, we fit the Gaussian perpendicular to the distribution, i.e., along the colour axis. Also, since the sequence is vertical, it is not affected by any variation in colour due to binarity of sources and hence it is safe to assume Gaussian distribution of sources. We then use the peak of the Gaussian curve as the mean locus of the PMS branch and restrict the membership to sources which lie within 3$\sigma$ from the mean. The dashed curve in Fig.~\ref{fig:fd_cmd} marks the mean locus of the PMS population and the continuous curves show the 3$\sigma$ limits.  At the brighter end, we limit the PMS selection at the saturation limit.  At the fainter end, we limit the source inclusion boundary at 90 \% completeness, where the PMS branch mostly merges with the field sequence. This limitation does not bias our results as we are interested only in the low-mass population (3 - 0.08 $M_{\sun}$) of the clusters.

Galactic interstellar extinction maps \citep{marshall2006} suggest that beyond the cluster NGC6611 at a distance between $\sim$2.8-3.5 kpc, there is an extinction jump from $A_k$ $\sim$0.6 to 1.4 mag, which is consistent with the location of the Scutum-Crux spiral arm towards the Galactic center \citep{vallee2008}. Using evolutionary tracks in J-H vs J CMD, \citet{oliveira2009} show that the dense population on the right of the PMS branch with redder colours (J-H $\sim$ 1.5 - 3 mag) are mainly background sources located in this spiral arm falling in the line of sight of NGC6611. This hypothesis is also explained in \citet{ngc6611guarcello2007} who shows that due to dust associated with the Eagle nebula in which the cluster is embedded, the background field stars are more reddened than the cluster stars. The J-H vs J CMD (see Appendix~\ref{sec:fd_CMD_of_all_clusters}) of NGC6611 shows this dense reddened background population. Due to the presence of this dense background population, the Gaussian peak of the PMS branch shifts towards the reddened sources at J-H $\sim$ 2.0 mag and below J = 16 mag. In order to obtain an unbiased locus of the PMS sources, we make a subset of the field decontaminated catalogue by excluding these background sources by giving a colour cut off of 1.5 mag in (J-H). We refit the Gaussian with the refined catalogue and follow the procedure mentioned above for field decontamination of NGC6611.

\subsection{Distance Estimation}
\label{sec:dist_est}

There have been various distances estimated for each cluster in the past (see Section~\ref{sample}). In order to constrain the distance  to each cluster uniformly, we use the data from Gaia Data Release 2 (\citealt{gaia2016,gaia2018}). Gaia data provides precise five-parameter astrometry (position, parallax and proper motion) for more than 1.3 billion sources in the Milky Way. Many studies have been carried out using the Gaia parallaxes to estimate the  distance to the star forming regions (e.g. \citealt{berlanas2019,herczeg2019}). \citet{kuhn2019} estimate the system parallax using the weighted median of individual stellar parallax measurements of Gaia and obtained the distances for NGC1893, NGC2362, NGC2244 and NGC6611 clusters. We adopt the distances estimated by \citet{kuhn2019} for these clusters (see Table~\ref{tab:phy_para}). Using the Gaia parallax and parameterised model inference approach, \citet{berlanas2019}  identify two different stellar groups superposed in the Cygnus OB2 association. They observe the main Cygnus OB2 group at $\sim1760$ pc and a foreground group at $\sim1350$ pc. We adopt the distance as 1760 pc for  Cygnus OB2 cluster for our analysis.

\begin{figure}
    \centering
    \includegraphics[width=\columnwidth]{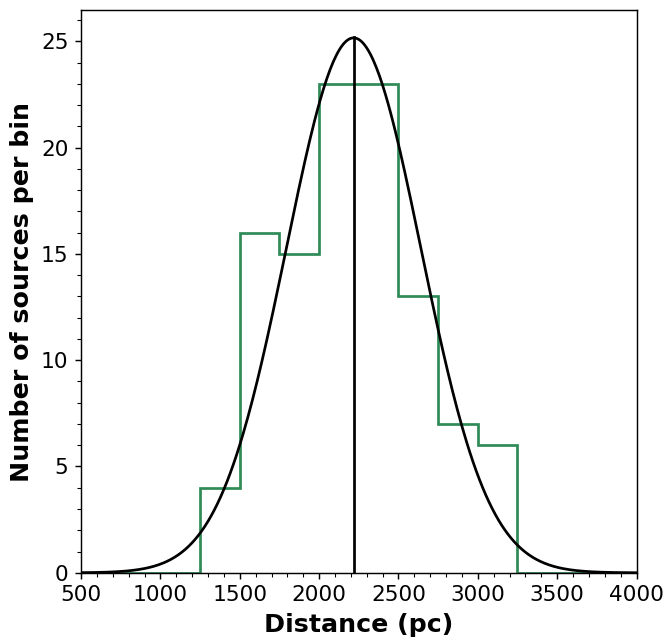}
    \caption{Histogram distribution of  distance of the candidate members in the sample cluster IC1848-West and the curve represents the Gaussian fit. The mean of this Gaussian fit is considered as the distance to the cluster.}
    \label{fig:dist_hist}
\end{figure}

For other clusters (i.e., IC1848-West, IC1848-East and Stock8) distance estimates using Gaia data is unavailable and we estimate the distances as follows. Using the Gaia parallax and a probabilistic approach, \citet{bailer2018} derive the distances to all the Gaia sources taking care of the non-linearity in parallax transformation. We obtain their distance catalogue for sources with parallax uncertainty $\leq 0.2$ within each  cluster radius (see section~\ref{sec:center&radius}). We cross-match this distance catalogue with the field decontaminated candidate PMS members (section~\ref{sec:fsd}) using a match radius of $1.2~\arcsec$ \citep{kuhn2019}. We model a Gaussian curve over the distance distribution of these sources. We converge the data to sources within  $1\sigma$ deviation from the peak (i.e. 68 \% confidence interval of the mean)   and the  mean and standard deviation of the refitted Gaussian distribution are considered as a proxy for the distance and its uncertainty, respectively. Fig.~\ref{fig:dist_hist} shows the histogram of the distance to the candidate PMS sources in the sample cluster IC1848-West for the converged data set along with Gaussian fitting. We cross-check our distance calculation by following the method described in \citet{kuhn2019}, for the above listed common clusters.  The values derived from both methods agree within 100 pc, which confirms the accuracy of our distance estimation. The distance thus used for individual regions are listed in Table~\ref{tab:phy_para}. Using the relation from \citet{xue2008} we have calculated the Galactocentric distance $(R_g)$ of the clusters and are given in Table~\ref{tab:phy_para}. 

\subsection{Extinction Estimation}
\label{sec:ext}

\begin{figure}
    \centering
    \includegraphics[width=\columnwidth]{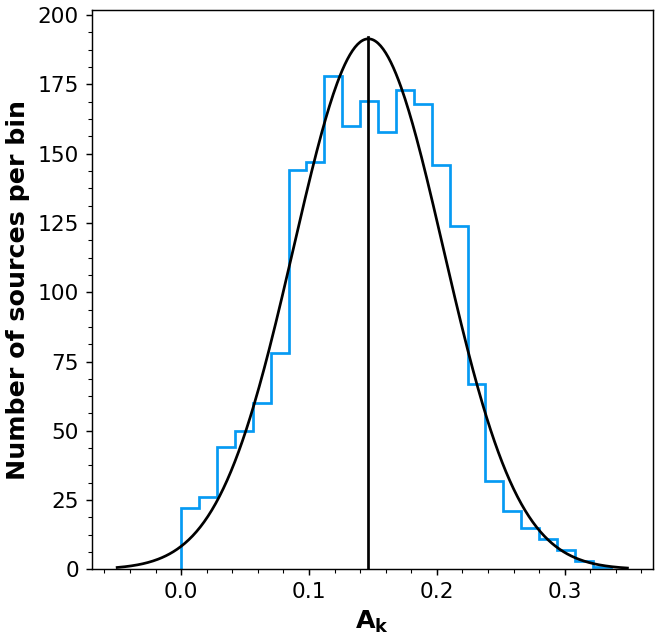}
    \caption{Histogram showing the extinction distribution of the candidate members in the sample cluster IC1848-West. The mean of the Gaussian fit is taken as the mean extinction value for the cluster.}
    \label{fig:ak_hist}
\end{figure}

There are two main factors that contribute to the extinction of stars in a cluster. One is the interstellar medium in the foreground of the cluster along the line of sight and the other is the localized parental molecular cloud in which the cluster is embedded. The extinction measurements of individual clusters by various studies are mentioned in section~\ref{sample}. In order to measure the extinction to the clusters uniformly, we follow the below method.  We use the field decontaminated candidate PMS  members  (section~\ref{sec:fsd}) for the extinction estimation of each cluster. We derive the K-band extinction towards the cluster using the extinction ratio $\frac{A_k}{A_v} = 0.114$ adopted from \citet{cardelli1989}, (i.e.,  $A_k = E(J-H)\times0.807$, where $E(J-H) = (J-H)_{obs}-(J-H)_{int}$).  The mean value of the (J-H) colours of  K and  M  type stars from \citet{pecaut2013} is taken as the intrinsic colour (i.e. $(J-H)_{int} = 0.6$ mag).  Since the spread in the intrinsic colors of low-mass objects are of the order of photometric uncertainty, this is a fair  approximation.  
Fig.~\ref{fig:ak_hist} shows the histogram distribution of $A_k$ of all the candidate PMS sources for the sample cluster IC1848-West.
The mean and standard deviation of the Gaussian fit are taken as the  extinction and uncertainty values, respectively for all the clusters.  We compare the above estimated mean extinction  with that of the extinction measured by dereddening to the intrinsic $(H-K)_0$ colour (ie, 0.2 mag) for the cluster IC1848-W and within uncertainty limit, both the values agree very well. The mean and errors of the extinction estimated  for each cluster are listed in Table~\ref{tab:phy_para}. Except Cygnus OB2, all the clusters have $A_K$ in the range of $\sim$ 0.05 - 0.3 mag (i.e., $A_V$ $<$ 2.5 mag)  with a narrow  spread of $\sim$ 0.07 mag, showing  that the reddening variation within the adopted radius  is relatively low in the clusters. The estimated extinction values of the clusters are used for  dereddening the sources to measure their age and mass. 

\subsection{Age}
\label{sec:age}

The age and age spread  of young  clusters are fundamental parameters which are among the most uncertain and difficult to constrain, especially at young ages (see \citealt{soderblom2014} for a review). Knowledge of these two timescales are critical for understanding the evolutionary appearance and state of a cluster and its star formation history \citep{lada2003}. Although we acknowledge that stars take different lengths of time to reach the ZAMS depending on their mass, in the simplest approximation, we picture all stars within a single cluster as having coalesced out of the interstellar medium at the same time. However, several studies have shown non-coeval stellar evolution of the young star-forming regions with a spread in the age estimates (e.g. \citealt{jessy2016,kraus2017,panwar2018} etc.). The classic method for estimating the age is through use of the Hertzsprung-Russell diagram (HRD), where the positions of member stars are compared with the locations of theoretical PMS evolutionary tracks and isochrones (e.g. \citealt{herczeg2015}). We use the online tool VO Sed Analyzer (VOSA)\footnote{\url{http://svo2.cab.inta-csic.es/theory/vosa/}} to obtain the luminosity and temperature of the candidate PMS members in the clusters. This tool builds Spectral Energy Distributions (SEDs) by using the J, H, K photometric data supplied along with the various online photometric catalogues in other wavelengths from 2MASS, FEPS, UKIDSS, WISE, VISTA, Spitzer, GLIMPSE, SDSS, Pan-Starrs, Dark Energy Survey, DECam and  VPHAS\footnote{\citep{2mass_vosa}, \citep{feps_vosa}, \citep{ukidss_vosa}, \citep{wise_vosa}, \citep{spitzer_vosa}, \citep{glimpse_vosa}, \citep{sdss_vosa}, \citep{panstars_vosa}, \citep{darkenergy_vosa}, \citep{vphas_vosa}} in the VO services, whenever available. We also input the distance and extinction information calculated in the previous sections for each cluster. 

\begin{figure}
    \centering
    \includegraphics[width=\columnwidth]{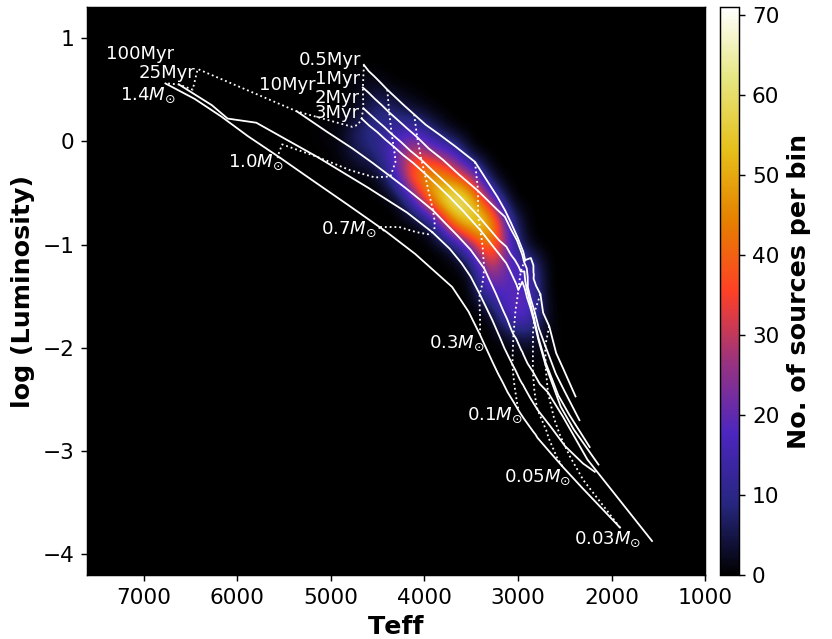}
    \caption{ H-R diagram of all the sources in the sample cluster IC1848-West with age $<$ 10 Myr. The luminosities and effective temperatures were obtained from the VOSA SED analysis. \citet{baraffe2015} evolutionary tracks and isochrones are shown as white dotted and continuous lines respectively.}
    \label{fig:age_hess}
\end{figure}

After correcting for the  distance and extinction values, VOSA compares the observed SED with synthetic photometry obtained from theoretical models for solar metallicity (BT-Settl models, \citealt{allard2014}). Using chi-square minimization technique we obtain the best fitting model and  corresponding physical parameters such as luminosity, effective temperature, age and mass for each source.  SED fitting however does not work for the faint sources in our catalog (e.g. J $\sim$ 17-18 mag) as most of them do not have counterparts in other surveys. The H-R diagram  of sources  is depicted as a hess diagram in Fig.~\ref{fig:age_hess}.  Evolutionary tracks and isochrones for various ages and masses from \citet{baraffe2015} are shown as white dotted and continuous lines respectively. Since the SED based membership analysis does not include a complete sample of members, the field decontaminated CMD based membership analysis, which is  more complete down to the low mass end, was preferred for the IMF estimation. The SED analysis  was used to obtain an average age of the member stars by fitting PMS isochrones of  \citet{baraffe2015}, as it gives more reliable age estimation compared to that of fitting isochrones directly on the NIR CMD.
 
\begin{figure}
    \centering
    \includegraphics[width=\columnwidth]{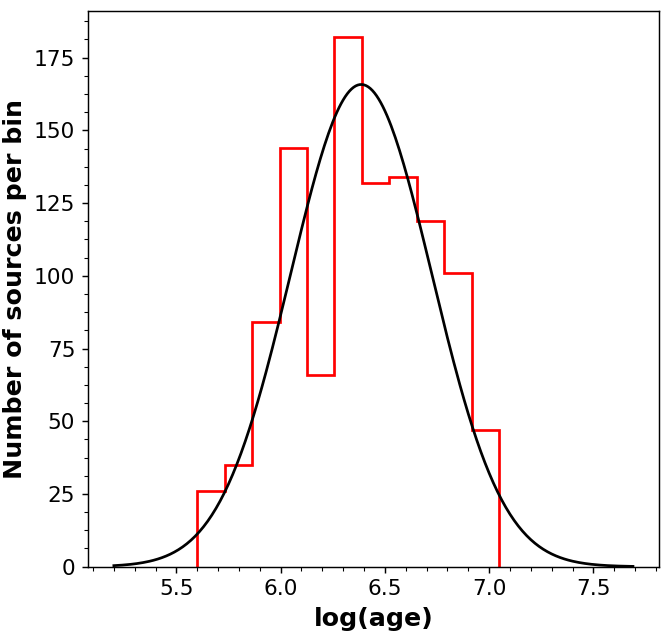}
    \caption{Histogram showing the age distribution of all the sources in the sample cluster IC1848-West with age $<$ 10 Myr. The mean of the Gaussian fit gives the mean age of the cluster.}
    \label{fig:age_hist}
\end{figure}

Since most of our sample clusters are associated with gas and dust (e.g., \citealt{sharma2007,koenig2008,oliveira2009}), their ages are likely to be less than 5 Myr \citep{leisawitz1989}. We consider  those sources with age less than 10 Myr for estimating the mean age of the cluster by fitting a Gaussian curve. Fig.~\ref{fig:age_hist} shows the age distribution of the sample cluster IC1848-West, fitted with the Gaussian curve. The mean age and age spread of the cluster is 2.2 $\pm$ 1 Myr.We rounded the mean age derived from the Gaussian fit to the nearest whole number and the estimated mean age and age spread of each cluster are listed in Table~\ref{tab:phy_para}. The mean age of our regions lie in the range $\sim$ 1-3 Myr, implying that the clusters are indeed young  and thus the effect of dynamical evolution is expected to be minimal. 

\begin{table*}
	\caption{Physical parameters of the clusters}
	\label{tab:phy_para}
    \begin{tabular}{|p{0.8in}p{0.6in}p{0.6in}p{0.6in}p{0.6in}p{0.6in}p{0.6in}p{0.6in}p{0.6in}|}               
		\hline
		\multirow{3}{*}[-3pt]{Cluster} & \multicolumn{4}{c}{Radius} & \multirow{2}{*}[-3pt]{Distance (d)} &  \multirow{2}{*}[-3pt]{Distance $(R_g)$} & \multirow{2}{*}[-3pt]{A$_k$} & \multirow{2}{*}[-3pt]{Median Age}\\
		\cmidrule{2-5}
		& Star count & Nearest neighbours & Mean & Mean & & & & \\
		& & & (arcmin) & (pc) & (pc) & (pc) & (mag)& (Myr)\\
		\hline
		IC1848-West & 4.2 & 3.8 & 4.0 & 2.6 & $2220\pm420$ $^b$ & $10080\pm350$ & $0.14\pm0.05$ & $2\pm1$\\
		IC1848-East & 2.9 & 3.1 & 3.0 & 2.0 & $2380\pm510$ $^b$ & $10230\pm430$ & $0.21\pm0.07$ & $2\pm1$\\
		NGC1893 & & & 3.0 $^a$ & 3.3 & $3790\pm600$ $^c$ & $12110\pm600$ & $0.15\pm0.07$ & $2\pm3$\\
		NGC2244 & 4.6 & 4.4 & 4.5 & 2.0 & $1550\pm95$ $^c$ & $9750\pm90$ & $0.10\pm0.06$ & $2\pm3$\\
		NGC2362 & 3.6 & 3.4 & 3.5 & 1.4 & $1332\pm70$ $^c$ & $9110\pm40$ & $0.04\pm0.03$ & $3\pm2$\\
		NGC6611 & 4.6 & 4.0 &4.3 & 2.2 & $1740\pm125$ $^c$ & $6690\pm120$ & $0.29\pm0.11$ & $2\pm1$\\
		Stock8 & 2.5 & 2.9 & 2.7 & 1.8 & $2290\pm460$ $^b$ & $10620\pm460$ & $0.15\pm0.07$ & $3\pm2$\\
		Cygnus OB2 & 2.8 & 3.2 & 3.0 & 1.5 & $1755\pm320$ $^d$ & $8220\pm10$ & $0.54\pm0.08$ & $2\pm2$\\
		\hline
	\multicolumn{5}{l}{$^a$\citep{sharma2007}}\\
	\multicolumn{5}{l}{$^b$This work}\\
	\multicolumn{5}{l}{$^c$\citep{kuhn2019}, mean of their error values are shown} \\
    \multicolumn{5}{l}{$^d$\citep{berlanas2019}, mean error value  is used}
	\end{tabular}
\end{table*}


\subsection{Mass-Magnitude relation}
\label{sec:mass-mag rel}

\begin{figure}
    \centering
    \includegraphics[width=\columnwidth]{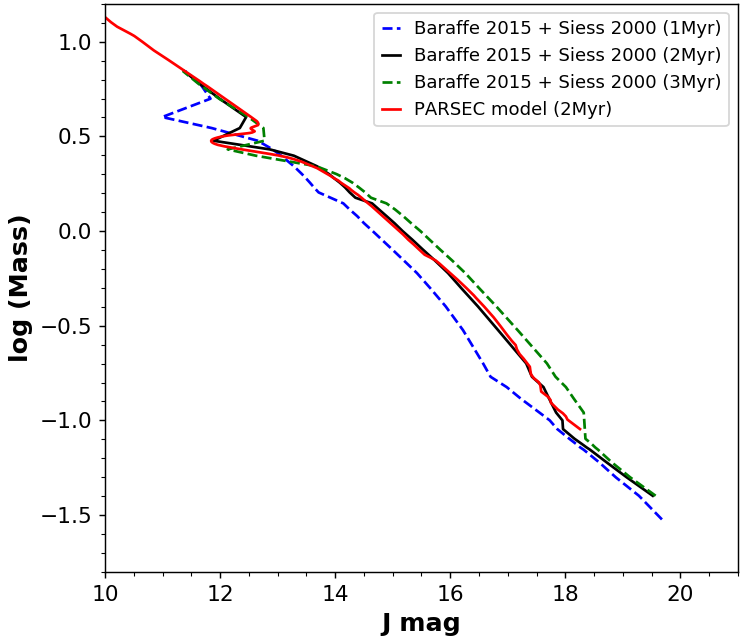}
    \caption{Mass-magnitude relation used for the IMF estimation of IC1848-West. The black solid curve is for the 2 Myr isochrones from \citet{baraffe2015} (for $<1.4M_{\sun}$) and \citet{siess2000} (for $>1.4M_{\sun}$), respectively and red curve is for PARSEC model of 2 Myr \citep{bressan2012}. The blue and green dashed curves are for 1 and 3 Myr from \citet{baraffe2015} and \citet{siess2000} isochrones, respectively.}
    \label{fig:mass-mag}
\end{figure}

In order to estimate the mass of the individual members of the clusters, we incorporate the PMS stellar evolutionary models. For the mass-magnitude conversion, we use the \citet{baraffe2015} models for sources with mass $<$ $1.4M_{\sun}$ and \citet{siess2000} models for mass $>$ $1.4M_{\sun}$, according to the respective cluster age. After correcting for the cluster distance and reddening, we convert the absolute magnitudes of the isochrones of respective age for each cluster to apparent magnitudes. By fitting a polynomial to both the models of above mass range, we obtain a relation between mass and magnitudes for J and K-bands. Using this relation, we estimate the mass of all the field decontaminated candidate members within each cluster in J and K-bands (see section~\ref{sec:fsd}). In Fig.~\ref{fig:mass-mag} we present the mass-magnitude relation for the 2 Myr old cluster IC1848-West in J band. Our final IMF is estimated using J and K-bands independently in order to compare our results and check for any biases in the analysis. 

The major sources of uncertainty in the above method to estimate the mass of PMS sources can be from the age spread and non-uniform reddening associated with the young clusters. However, for younger ages ($<$5 Myr), the PMS isochrones in J vs J-H or J vs H-K planes are almost vertical (see \citealt{kuhn2017,jessy2017}). This is because the colour change (in J--H or H--K ) of the low-mass PMS sources  due to age variations within the range of $\sim$ 2 - 5 Myr is negligible (see \citealt{siess2000,pecaut2013,baraffe2015}). Since the reddening variation within the clusters in our list is minimal (see Table~\ref{tab:phy_para}), the above assumptions of mean age and mean reddening of the clusters are reasonably valid to estimate the mass-magnitude relation of the PMS sources. In order to check any uncertainty associated  with possible age spread of the clusters in mass estimates (see \citealt{neichel2015}), we over plot the isochrones of age  1 and 3 Myr from \citet{siess2000} and \citet{baraffe2015} in Fig.~\ref{fig:mass-mag}. The uncertainty in the mass-magnitude relation for $\sim$2 Myr age spread is of the order of the size of the mass bin considered for IMF calculation (see section~\ref{imf}) and hence we ignore this effect.
Another form of uncertainty in the mass estimates of low-mass PMS members is from the lack of consistency among different stellar evolutionary models (e.g. \citealt{soderblom2014,herczeg2015} and references therein). In Fig.~\ref{fig:mass-mag}, we over plot the mass-magnitude curve for the PMS isochrone of 2 Myr age from PARSEC evolutionary models \citep{bressan2012}. It is evident that the mass-magnitude relation of PARSEC models matches with that of the  relation constructed from the 2 Myr isochrones from \citet{siess2000} and \citet{baraffe2015} and  hence we ignore the effect of discrepancies among various PMS models in our analysis.

\subsection{Initial Mass Function}
\label{imf}

In this section we discuss the distribution of IMFs in our study.   Although observations of embedded clusters reveal the presence of binary systems within them that are difficult to resolve, as discussed in section~\ref{introduction}, the effect of binarity on IMF estimates is relatively small (\citealt{harayama2008,zeidler2017,muzic2017,suarez2019}) and hence  we do not account for  any unresolved companions in our analysis. The  system IMF treats binaries and multiple systems as single stars rather than multiple stars and hence is ideal to compare with our observations. Thus the IMFs  we derive are the system IMFs \citep{lada2003}. We obtain the IMF for the field decontaminated PMS sources (section~\ref{sec:fsd}) and within 90 \% photometric completeness (see Table~\ref{tab:cluster_list}), which are mostly in the magnitude range of $\sim$ 13-18 mag (see Fig.~\ref{fig:fd_cmd} and Appendix~\ref{sec:fd_CMD_of_all_clusters}) and mass range of $\sim$ 3 - 0.08 M$_{\sun}$ (see Fig.~\ref{fig:mass-mag}). We calculate the system IMF by counting the number of stars in a logarithmic mass interval of bin size, log(m) = 0.2. 

Pioneering work done by \citet{salpeter1955} introduced a power-law for the IMF of the form,
\begin{equation}
    \phi(log m) = \frac{dN}{dlog m} \propto m^{-\Gamma}
\end{equation}

where m is the mass of a star, N is the number of stars in a logarithmic mass range log m + dlog m and $\Gamma$ was found to be $\sim1.35$ which is generally referred  as the Salpeter slope. Later it was recognized that the IMF was probably not a single power-law over all stellar masses. \citet{kroupa2001} presents a multi-segment power law, where the slope of the IMF at lower masses was found to be shallower than the Salpeter slope at higher masses. The logarithmic formalism of the IMF by  \citet{chabrier2003} is also been widely used as it provides a description of the IMF as a log-normal function at the low-mass end and power-law form above 1 M$_{\sun}$.

In this analysis we describe the derived system IMF using the log-normal distribution, i.e.,
\begin{equation}
    \xi(log m) \propto e^{-\frac{(log m-logm_c)^2}{2\sigma^2}}
\end{equation}

where $m_c$ is the characteristic mass (the mass at the peak of the distribution) and $\sigma$ is the standard deviation.

\begin{figure*}
    \centering
    \includegraphics[width=\textwidth]{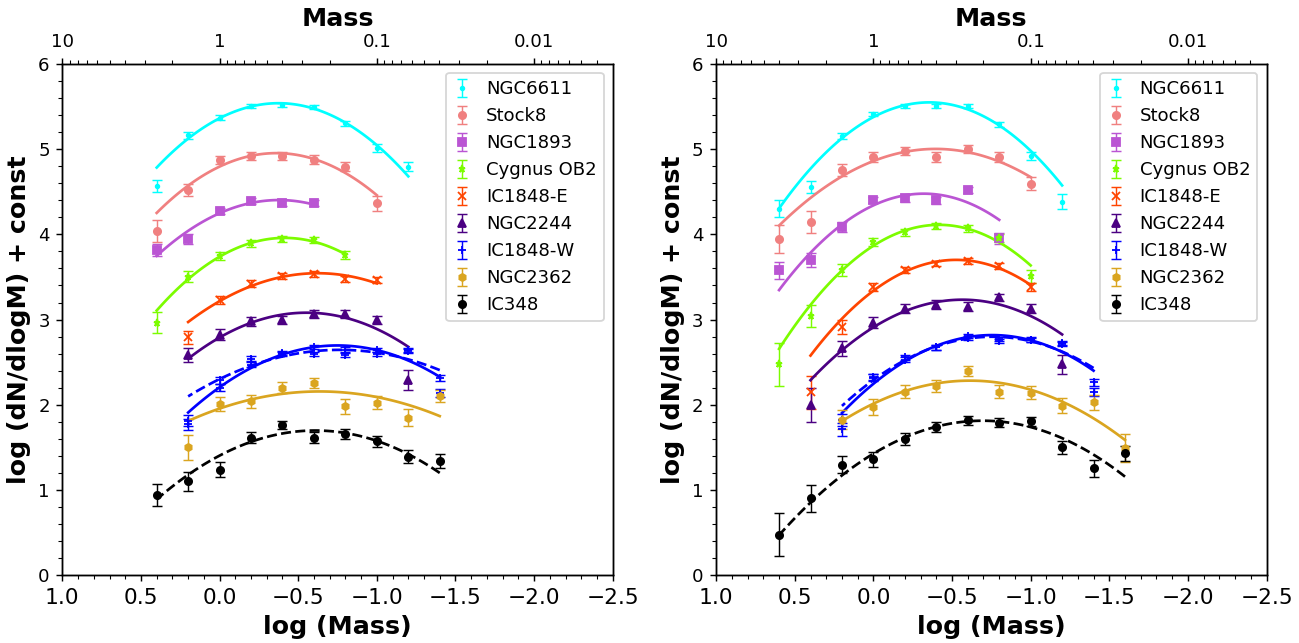}
    \caption{The IMF derived from the J-band (left panel) and K-band (right panel) for all the clusters. Error bars denote the Poisson error on each data point. The curves are the log-normal fits to the individual cluster IMF. The blue dotted curve shows the IMF estimated by applying extinction correction to individual sources of IC1848-West cluster (refer text).}
    \label{fig:individual_imf}
\end{figure*}

Fig.~\ref{fig:individual_imf} shows the individual IMFs of all the clusters obtained from the mass estimated from J band (left panel) and from K band (right panel). The error bars represent the associated Poisson errors and the continuous curves are the log-normal fits to the individual cluster IMFs. The respective characteristic mass $m_c$ and $\sigma$ (the spread in the log-normal distribution) are listed in Table~\ref{tab:IMF_details} for each cluster. The peak mass of the individual cluster lies in the range of $\sim$ 0.18 - 0.48 $M_{\sun}$ and $\sigma$ is in the range of $\sim$ 0.39 - 0.66. 

We compare our results with a nearby young cluster IC348, which is one of the well studied star forming regions and has well characterized membership analysis by several studies. IC 348 resides in the Perseus molecular cloud and is one of the nearest ($\sim320$ pc; \citealt{ortiz2018}) and richest star forming regions with an age of $\sim$ 2-3 Myr \citep{lada2006}. \citet{luhman2016} presents a complete census of members of IC348 using optical and NIR spectral analysis. The data is nearly complete down to  K $<$ 16.8 mag for $A_J$ $<$ 1.5 mag, which corresponds to mass limit of $\sim$ 0.01 $M_{\sun}$ for an age of $\sim$ 3 Myr. For comparison, we estimate the IMF of IC348 by the same method as that of our clusters in this study. Using the source list from \citet{luhman2016} and mass-magnitude relation as mentioned in the previous section, we estimate the mass of the cluster members within IC348. The IMF thus obtained for IC348 is plotted along with our clusters in Fig.~\ref{fig:individual_imf}. The general form of IMF and the values of $m_c$ (0.25 $M_{\sun}$) and $\sigma$ (0.50) of IC348 are well  consistent with the clusters in our list.

The mean values of $m_c$ and $\sigma$ for the IMF estimated from J band are $0.32\pm0.02$ $M_{\sun}$ and $0.49\pm0.02$ and from  K band are $0.32\pm0.01$ $M_{\sun}$ and $0.45\pm0.02$, respectively. In order to assess the effect of binning, the above analysis was repeated by shifting the mass bin by log(m) = 0.1 $M_{\sun}$ as well as by varying the bin sizes. After applying these variations, the difference in the estimated characteristic mass, $m_c$ and $\sigma$ was found to be agreeing within the uncertainty range for each cluster. As discussed in sections ~\ref{sec:fsd} and ~\ref{sec:ext} (also refer Appendix~\ref{sec:fd_CMD_of_all_clusters}) the clusters exhibit minimal differential extinction within the area considered for the study facilitating the use of an average extinction value in deriving the IMF. Nevertheless to eliminate any uncertainties that might arise due to the consideration of uniform extinction in a cluster region, extinction correction was applied to individual sources in the sample cluster IC1848-West to obtain their absolute magnitudes. These absolute magnitudes were then compared to the PMS stellar evolutionary models to obtain their masses. With the newly obtained masses, the IMF of IC1848-West cluster was re-estimated as shown in Fig.~\ref{fig:individual_imf} as a dashed curve. The values of $m_c$ and $\sigma$ estimated from J band are $0.17\pm0.07$ $M_{\sun}$ and $0.61\pm0.09$ and from K band are $0.18\pm0.04$ $M_{\sun}$ and $0.50\pm0.04$ respectively. The values of the characteristic mass ($m_c$) and $\sigma$ obtained by applying extinction correction to individual  sources are in agreement with the values estimated by applying average extinction to the cluster.

\begin{figure*}
    \centering
    \includegraphics[width=\textwidth]{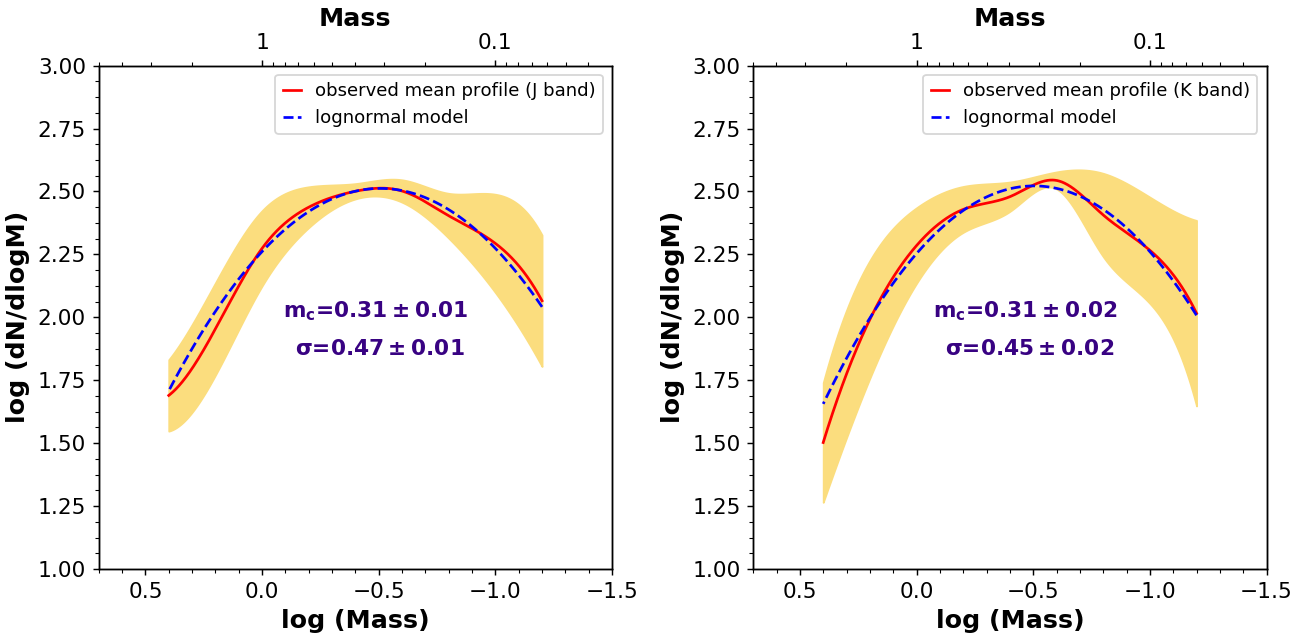}
    \caption{Mean IMF of all the 8 clusters in the study including IC348 (red curve) derived using J-band mag (left panel) and K-band mag (right panel). Blue dotted curve is the log-normal fit to the distribution. The shaded region marks the 1 sigma deviation from the mean. }
    \label{fig:mean_imf}
\end{figure*}

We also obtain the mean values of each mass bin in Fig.~\ref{fig:individual_imf} after normalizing the peak value of IMF to one of the clusters, i.e., IC1848-West, and obtain an empirical mean IMF of all the 8 clusters in this study along with IC348 and is shown in Fig.~\ref{fig:mean_imf} (red curve). The shaded region marks the 1 sigma deviation from the mean value in each mass bin. The IMF generally flattens out between $0.2 - 0.7M_{\sun}$ and then drops down on both sides, which is in agreement with the general form of IMF of young clusters (\citealt{neichel2015,maia2016,moraux2016,jessy2017,suarez2019} and references therein). The log-normal fit to the above mean distribution (blue dashed curve) gives $m_c = 0.31\pm0.01$ $M_{\sun}$ and $\sigma = 0.47\pm0.01$  for J band and $m_c = 0.31\pm0.02$ $M_{\sun}$ and $\sigma = 0.45\pm0.02$  for K band. The mean IMF distribution of the clusters in this study correlates well with that of the Galactic field mass function which has a characteristic mass $m_c$ = 0.25 $M_{\sun}$ and $\sigma$ = 0.55 \citep{chabrier2003}.

As discussed in section~\ref{sec:center&radius}, we have restricted our individual cluster radii to 3 sigma above the mean density of the background region in order to limit the differential reddening  as well as field star contaminants in a larger area. As a result, there are chances of exclusion of a few low-mass members lying beyond this radius for IMF estimates (e.g. due to primordial mass segregation,  \citealt{andersen2017,kaur2020}). In order to test this, we estimate the IMF of a sample cluster, Stock 8, for a  radius above the background  density, i.e., 5$\arcmin$, which is $\sim$ 2.3 $\arcmin$ larger than the radius considered previously. We repeated the same process for this larger radius to obtain the IMF and the values for $m_c$ and $\sigma$ estimated using the J-band are 0.36 $\pm$ 0.04 $M_{\sun}$ and 0.47 $\pm$ 0.04, respectively. Within uncertainties, the $m_c$ and $\sigma$ values for larger radius are consistent with the values estimated for smaller radius. Therefore we consider that the effect of stars lying beyond the 3 sigma radius of the clusters on the overall shape of the IMF is unlikely to be significant. 

Before we discuss the implications of our results, it is important to report some of the earlier IMF studies on these clusters. \citet{jessy2017} derived the IMF of Stock 8 using optical and deep IR data for a cluster radius of 3$\arcmin$. Their log-normal distribution to the low-mass end (0.08-1 $M_{\sun}$) yielded a peak mass of $m_c$=0.43$M_{\sun}$, which is consistent with our results. \citet{muzic2019} studied the low-mass part of the IMF of NGC2244 represented by two power-laws. Although a direct comparison with our results is difficult, it is worth mentioning that \citet{muzic2019} reported that there is no effect due to the lack/presence of massive OB stars on the formation efficiency of low-mass stars. 

\subsection{Monte Carlo simulations to assess various uncertainties}
\label{subsec:MCmethod}

We have also assessed the impact of possible systematic bias on the IMF caused by the  uncertainties associated with the observational parameters such as distance, age and extinction of the regions as well as the  evolutionary models used. This assessment was done for all the clusters by employing the Monte Carlo (MC) simulations using the field decontaminated sources by independently generating random values for the three parameters - age, distance and extinction. For each of the three parameters we randomly generated 2000 values normally distributed within a range corresponding to their respective mean with dispersion equal to their errors listed in Table~\ref{tab:phy_para}.  Following which 1000 iterations were processed wherein each iteration used a randomly picked set of values pertaining to the three parameters and the mass corresponding to J-band was estimated utilizing the same evolutionary models mentioned in section ~\ref{sec:mass-mag rel} (i.e., \citet{baraffe2015} models for sources with mass $<$ $1.4M_{\sun}$ and \citet{siess2000} models for mass $>$ $1.4M_{\sun}$). Thus using the mass obtained through each iteration, the respective IMF was constructed by fitting a log-normal function and $m_c$ and $\sigma$ were estimated. The average values of $m_c$ and $\sigma$ from the 1000 iterations for each cluster are tabulated in Table~\ref{tab:IMF_details}. The mean values of $m_c$ and $\sigma$ for the eight clusters are 0.26$\pm$0.04 and 0.43$\pm$0.08, respectively. For majority of the clusters, the values obtained from the simulations are within 3 sigma uncertainty of the values obtained from the fitting method above (refer Section \ref{imf}). Also, in order to account for any possible bias associated with the use of \citet{baraffe2015} models in the above IMF estimates of the clusters, we have done the MC simulations detailed above using a different model set (i.e.,  PARSEC evolutionary models; \citealt{bressan2012}) for a sample cluster IC1848-West. The use of a different evolutionary model resulted in a marginal shift in the mean values of $m_c$ and $\sigma$ with $m_c$=0.27$\pm$0.13 and $\sigma$=0.41$\pm$0.28, which are within the average values obtained for the eight clusters within 3 sigma listed in Table ~\ref{tab:IMF_details}.

\begin{table*}
    \centering
	\caption{Characteristic mass (m$_c$) and $\sigma$ values from Fig.~\ref{fig:individual_imf}}
		\begin{tabular}{p{3.5cm}p{1.8cm}p{1.8cm}p{1.8cm}p{1.8cm}p{1.8cm}p{1.8cm}}
		\hline
		\multirow{4}{*}{Cluster} & \multicolumn{4}{c}{$^a$From fit} & \multicolumn{2}{c}{$^b$From MC simulation}\\
		\cmidrule{2-7}
		& \multicolumn{2}{c}{J band} & \multicolumn{2}{c}{K band} & \multicolumn{2}{c}{J band}\\
    	\cmidrule{2-7}
		& $m_c$ & $\sigma$ & $m_c$ & $\sigma$ &  $m_c$ & $\sigma$\\
		& ($M_{\sun}$) &  & ($M_{\sun}$) & & ($M_{\sun}$) &\\
		\hline
		IC1848-West & $0.18\pm0.04$ & $0.50\pm0.05$ & $0.18\pm0.04$ & $0.47\pm0.04$ & $0.15\pm0.10$ & $0.53\pm0.26$ \\
 		IC1848-East & $0.24\pm0.04$ & $0.51\pm0.05$ & $0.30\pm0.02$ & $0.41\pm0.02$ & $0.24\pm0.11$ & $0.40\pm0.24$\\
 		NGC1893 & $0.42\pm0.05$ & $0.45\pm0.06$ & $0.48\pm0.06$ & $0.40\pm0.07$ & $0.38\pm0.15$ & $0.37\pm0.22$\\
 		NGC2244 & $0.28\pm0.05$ & $0.48\pm0.06$ & $0.27\pm0.05$ & $0.46\pm0.06$ & $0.23\pm0.11$ & $0.43\pm0.22$\\
 		NGC2362 & $0.23\pm0.13$ & $0.66\pm0.18$ & $0.24\pm0.06$ & $0.55\pm0.07$ & $0.13\pm0.05$ & $0.46\pm0.26$\\
 		NGC6611 & $0.42\pm0.01$ & $0.42\pm0.02$ & $0.44\pm0.02$ & $0.40\pm0.02$ & $0.42\pm0.15$ & $0.40\pm0.15$\\
 		Stock8 & $0.44\pm0.03$ & $0.42\pm0.04$ & $0.40\pm0.05$ & $0.49\pm0.05$ & $0.21\pm0.14$ & $0.44\pm0.40$\\
 		Cygnus OB2 & $0.39\pm0.01$ & $0.41\pm0.02$ & $0.38\pm0.02$ & $0.39\pm0.02$ & $0.31\pm0.14$ & $0.40\pm0.18$\\
 		IC348 & $0.25\pm0.05$ & $0.52\pm0.06$ & $0.20\pm0.04$ & $0.52\pm0.04$ & $^c$- & - \\
		\hline
		Mean & $0.32\pm0.02$ & $0.49\pm0.02$ & $0.32\pm0.01$ & $0.45\pm0.02$ & $0.26\pm0.04$ & $0.43\pm0.08$\\
		\hline
		$^*$Mean & $0.31\pm0.01$ & $0.47\pm0.01$ & $0.31\pm0.02$ & $0.45\pm0.02$ & & \\
		\hline
	\multicolumn{7}{l}{$^a$Values obtained by fitting the individual IMF of the clusters, shown in Fig.~\ref{fig:individual_imf}.}\\
	\multicolumn{7}{l}{$^b$Values obtained by fitting the IMF after accounting for the various uncertainties through Monte Carlo simulations (refer Section~\ref{subsec:MCmethod}).}\\
	\multicolumn{7}{l}{$^c$For IC348, Monte Carlo simulation was not performed since physical parameters were taken from \citet{luhman2016}.}\\
	\multicolumn{7}{l}{$^*$Values obtained after fitting the mean IMF of all the clusters, shown in Fig.~\ref{fig:mean_imf}.}
	\label{tab:IMF_details}
	\end{tabular}
\end{table*}

\begin{table}
	\caption{Mean log(Mass) and log(dN/dlogM) values used for  Fig.~\ref{fig:mean_imf} }
	\begin{tabular}{lcccccc}
		\hline
		\multicolumn{2}{c}{J band} & \multicolumn{2}{c}{K band}\\
		\cmidrule{1-4}
		log(Mass) & log(dN/dlogM) & log(Mass) & log(dN/dlogM)\\
		\hline
		-1.2 & 2.07 & -1.2 & 2.02\\
		-1.0 & 2.29 & -1.0 & 2.27\\
		-0.8 & 2.40 & -0.8 & 2.41\\ 
		-0.6 & 2.50 & -0.6 & 2.54\\
		-0.4 & 2.50 & -0.4 & 2.48\\
		-0.2 & 2.44 & -0.2 & 2.43\\
		 0.0 & 2.27 & 0.0 & 2.28\\
		 0.2 & 1.95 & 0.2 & 1.99\\
		 0.4 & 1.69 & 0.4 & 1.50\\
		\hline
	\label{tab:Mean mass values from IMF}
	\end{tabular}
\end{table}

\section{Discussion}
\label{discussion}

Embedded clusters are the basic units of star formation and their study can address some fundamental astrophysical problems like cluster formation and evolution through the form and universality of the stellar IMF. The embedded phase of cluster evolution appears to last between 2 to 5  Myrs, and clusters with age greater than 10 Myrs are rarely associated with molecular gas (\citealt{lada2003,leisawitz1989}). Several  studies have been carried out to understand the nature of the IMF, especially the low-mass regime has been the subject of numerous observational and theoretical studies over the past decade (see \citealt{offner2014} for a review).

\subsection{Characteristic stellar mass and theoretical implications}

The high mass stars generally follow the Salpeter mass function \citep{salpeter1955}. However, the IMF is not so well constrained at lower masses and appears to flatten below $1M_{\sun}$ with  a turnover between $0.1 - 0.7M_{\sun}$ \citep{chabrier2003}. The stellar mass at which this  transition in the slope occurs is considered to depend generally on the physical properties of the underlying molecular cloud (\citealt{larson2005,elmegreen2008}). The peak of the IMF is a key constraint for star formation models since it is not a scale-free parameter (unlike the high-mass slope) and thus additional physics beyond gravity-driven accretion or turbulence is required to set it \citep{krumholz2014}. Possibilities include the thermal Jeans Mass (e.g. \citealt{larson2005}), the turbulent Jeans Mass (e.g. \citealt{hennebelle2008}), radiative feedback (e.g. \citealt{bate2009}) and initial cloud density (e.g \citealt{jones2018}). Predictions for how the IMF  and in particular the peak mass of the IMF behaves in different environments, changes depending on which of these processes dominate. Simulation studies  by \citet{krumholz2016} suggest that radiative heating is the main driving mechanism of the characteristic mass of IMF. This study shows that the efficiency of cloud  fragmentation  reduces as radiative heating increases, which eventually leads to a top-heavy IMF.  On the other hand, \citet{conroy2012} suggests that the  radiative ambient pressure plays a significant role  for generating bottom-heavy IMFs (at increasing pressure) as observed in elliptical galaxies with a history of starburst-generating mergers.  Additional kinetic feedback such as   stellar winds, protostellar outflows/jets and UV ionization etc. are also likely  to affect the efficiency of star formation (e.g. \citealt{li2006}). However, it is still a matter of debate how and if they ultimately affect stellar mass distribution, i.e., the IMF. Thus, exploring various star forming environments is a valuable tool for understanding the standard form of IMF (see \citealt{hosek2019}).

In the standard picture of the star formation process, beginning from the compression of gas in a GMC leading to the collapse and formation of protostars and their evolution into PMS objects through accretion, different environmental conditions may play a role in shaping the final products like IMF (\citealt{prisinzano2011} and references therein). In this process, it has been shown that massive stars affect the evolution of their natal molecular clouds through their strong stellar winds and UV radiation (\citealt{murray2010,dale2012,walch2013,reyraposo2017,kim2019}). The ionizing UV radiation has strong competing effects on surrounding molecular cloud and subsequent star formation (eg. \citealt{dale2017,gavagnin2017,geen2017,kim2018,kruijszsen2019}). One is negative feedback on the star formation activity that disperses the remaining molecular cloud and truncates further star formation. The other is positive feedback as the interaction triggers new episodes of star formation (eg. \citealt{deharveng2012,jessy2013,samal2014,jessy2016,panwar2019}). \citet{krumholz2016} suggests that radiation feedback is the key process in determining the location of the peak of the IMF.  Considering that the clusters in this study are under diverse radiation environments (see below), it is therefore useful  for  understanding the role of UV radiation feedback  on the form of the IMF. 

\subsection{IMF under diverse environmental conditions} 

\begin{figure}
    \includegraphics[scale=0.54]{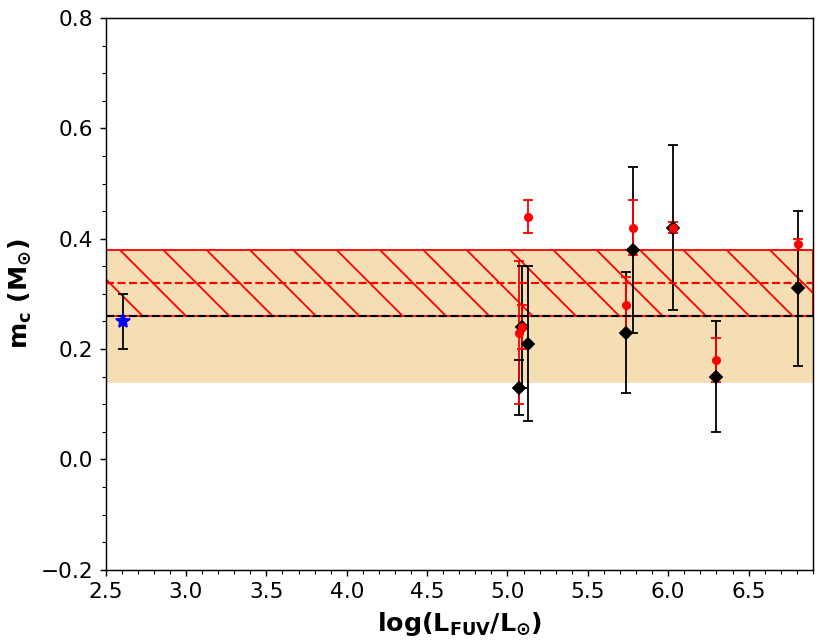}
    \includegraphics[scale=0.54]{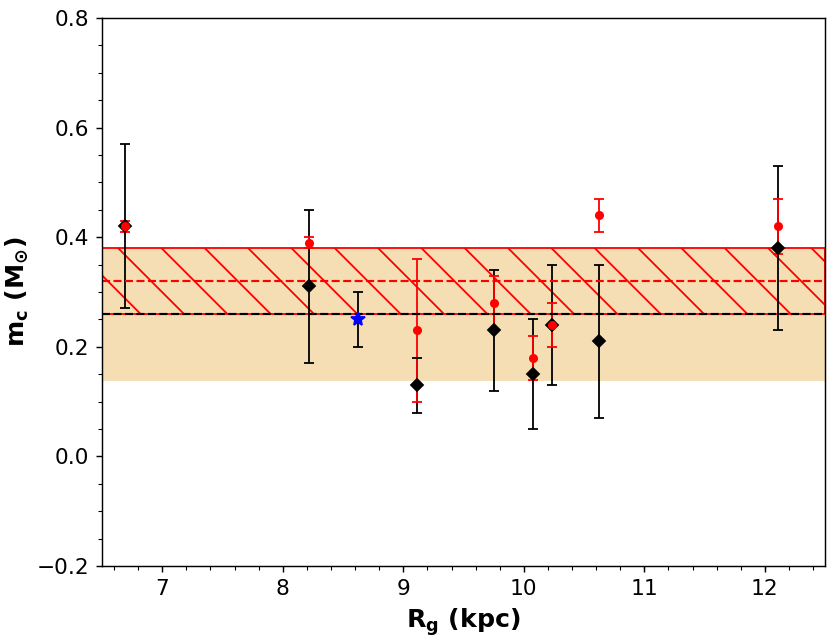}
    \includegraphics[scale=0.54]{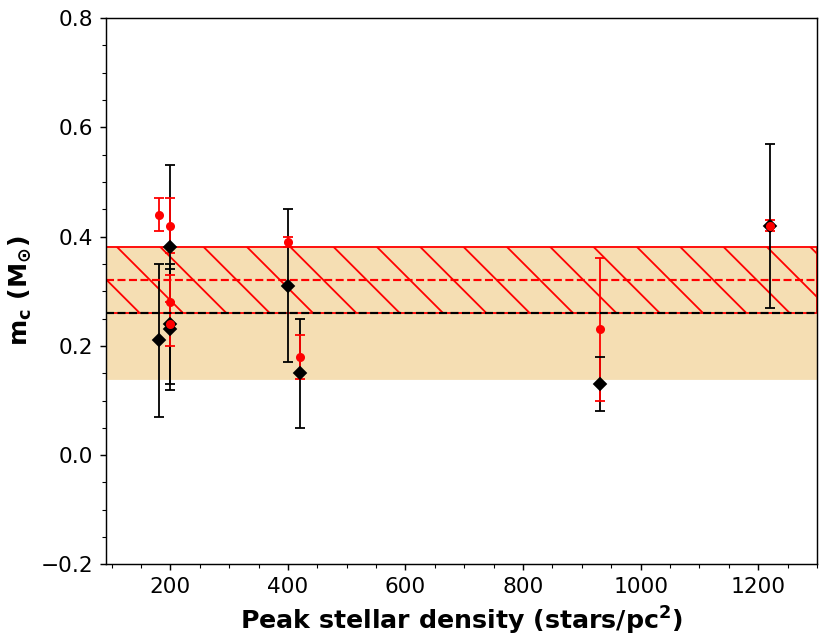}
    \caption{The distribution of characteristic stellar  mass ($m_c$) as a function of  $L_{FUV}$ {\it (top)}, as a function of Galactocentric distance ($R_g$) {\it (middle)}, and as a function of peak stellar density {\it (bottom)} for the clusters obtained from Fig~\ref{fig:individual_imf} (red dots) and from Monte Carlo simulations (black dots). The blue source represents the nearby low-mass cluster IC348. The red dashed line and hatched region represent the mean value of $m_c$ = 0.32 $M_{\sun}$ and 3 sigma deviation respectively from Fig~\ref{fig:individual_imf}. The black dashed line and yellow shaded region mark the mean value of $m_c$ = 0.26 $M_{\sun}$ and 3 sigma deviation respectively, after accounting for the various sources of uncertainties using the Monte Carlo simulations.}
    \label{fig:uv_rg_comparison}
    \end{figure}

We have carried out this study to test the universality of the IMF in low-mass end down to brown-dwarf limit for star-forming regions of diverse environmental conditions in terms of UV radiation, Galactic location, stellar density etc. We estimate the UV radiation field strength,   $log(L_{FUV}/L_{\sun})$ and $log(L_{EUV})$   from the respective massive stars present in each cluster listed in  section ~\ref{sample}.  
UV luminosities corresponding to the spectral types are obtained  from \citet{guarcello2016} and  \citet{thompson1984}, and we add them for all the massive stars  present in a given cluster. The eight clusters selected in this study are embedded in massive stellar environments of radiation field strength $log(L_{FUV}/L_{\sun})$ $\sim$ 5.07 to 6.81 and  $log(L_{EUV})$  $\sim$ 48.8 to 50.85 photons/s, whereas, for IC348, these values are $\sim$ 2.6 and $\sim$ 42.2 photons/s, respectively. These clusters are located at a heliocentric distance of $\sim$1-4 kpc, ($R_g$ $\sim$6-12 kpc) and the associated molecular cloud mass ranges from  $\sim$10$^4$ - 10$^5$ $M_{\sun}$ (see references in section~\ref{sample}). The  peak stellar density at the cluster centers varies between $\sim$170 - 1220 stars/pc$^2$. The radiation strength, stellar density, cloud density, Galactocentric distance  etc. vary by several orders among the clusters and hence these regions can be considered as in diverse environmental conditions. In Fig.~\ref{fig:uv_rg_comparison} (top panel), we compare the characteristic mass distribution as a function of $L_{FUV}$ for the nine clusters. We do not find any strong  dependence of radiation field strength on the shape of IMF (see Fig.~\ref{fig:individual_imf}) as well as on characteristic stellar  mass (see Fig.~\ref{fig:uv_rg_comparison})  as a function of above parameter. 

In the disc of our Galaxy, a number of ISM properties are found to be varying as a function of Galactocentric distance \citep{rigby2019} such as the metallicity \citep{caputo2001,luck2011}, molecular-to-atomic gas ratio (e.g., \citealt{sofue2016}), interstellar radiation field (e.g., \citealt{popescu2017}) and dust temperature (e.g., \citealt{urquhart2018}).  However, the outcome of various star formation activities such as the clump-formation efficiency (or the dense-gas mass fraction), star-formation efficiency or the physical properties of the molecular clumps do not seem to vary as a function of Galactocentric radius (see \citealt{moore2012,eden2013,eden2015,rigby2019} and references therein).  Similarly, one of the main outcome of star formation activity, IMF, needs to be tested across various Galactocentric distances. In Fig. \ref{fig:uv_rg_comparison} (middle panel), we plot the characteristic stellar mass as a function of the Galactocentric radius and we do not find any systematic trend in $m_c$ as a function $R_g$. Also, the form of low-mass end of IMF do not vary with $R_g$ (see Fig.~\ref{fig:individual_imf}).  Similarly, in Fig. \ref{fig:uv_rg_comparison} (bottom panel) it is seen that the form of IMF or characteristic stellar mass do not show any systematic variation as a function of stellar density for the various clusters under this study. In summary, we do not observe any systematic variation in our analysis, implying no strong environmental effects on the clusters under this study. 


\subsection{Comparison with other  regions} 

As discussed earlier, the behaviour of IMF in diverse environmental conditions are yet to be well understood. Some of the young star forming regions in the solar neighbourhood whose IMF down to the sub-solar regime have been extensively studied are discussed here. Among them, the Orion Nebula Cluster (ONC) located at $\sim$ 470pc is reported to have a log-normal distribution in the low-mass end ($\sim 0.02 - 3M_{\sun}$) of the IMF with the characteristic mass of 0.28$\pm$0.02$M_{\sun}$ and $\sigma =$ 0.38$\pm$0.01 \citep{dario2012}. Similarly  NIR survey of the nearby star forming region, $\sigma$ Orionis, by   \citet{pena2012}  shows that the low-mass end of the IMF has a similar distribution to that of other nearby young clusters, with the characteristic mass of 0.27$\pm$0.07 and $\sigma = 0.57\pm0.11$ in the mass range 0.006-19 $M_{\sun}$. Another such star-forming region is the 25 Orionis (25 Ori) located at a distance of $\sim$350pc and age $\sim$ 7-10Myr. \citet{suarez2019} described the IMF of 25 Ori using various functional forms for different cluster radius.  The log-normal fit to the derived IMF in the mass range of 0.01-13$M_{\sun}$ for an area of 0.5$^\circ$ yields the values of $m_c$ and $\sigma$ as $0.31\pm0.06$ and $0.51\pm0.08$, respectively. Here we would like to note that within uncertainties, the IMFs of all these nearby star forming regions in the solar neighbourhood  correlate well with our results of 8 young clusters, which are relatively distant and with diverse physical parameters and environmental conditions. 

Likewise we also compare our results with some of the most massive young Galactic star clusters like Westerlund 1, NGC3603 and RCW 38. Westerlund 1 (Wd1) is a super massive cluster located at a distance of $\sim$ 3.7-5 kpc and of age $\sim$ 3-5 Myr old. Using deep HST data, \citet{andersen2017} derived the mass function of the supermassive cluster as a function of radius. They show that the flattening at the low mass end of the IMF is similar to that of nearby low mass star-forming regions and a log-normal fit to the IMF shows the width of the distribution of the sub-solar population to be comparable  or slightly less  than that of the Galactic field ($\sigma$ $\sim$ 0.33-0.44). Another massive region is the luminous, optically visible compact cluster NGC3603 located at a distance of $\sim$ 7kpc \citep{pandey2000} in the Carina arm of our Galaxy. The core of NGC3603 contains a Trapezium-like system (HD 97950) with about 50 massive stars which makes it analogous to the core of R136 in 30 Doradus in the Large Magellanic Cloud. Hence NGC3603 is considered as a local template of starburst regions in distant galaxies \citep{nurnberger2002}. A direct comparison of our results with the IMF of NGC3603 is not feasible as most of the studies derive the power-law distribution of the IMF. Nevertheless, we report the observations of \citet{ram2001} which put forward the idea that low-mass IMF of NGC3603 and other similar young clusters have no dependence on Galactic longitude, $R_g$ and cluster age. RCW38 is another young ($\sim$1 Myr) massive dense star forming region at a distance of $\sim$1.7 kpc \citep{wolk2006}. \citet{muzic2017} studied the low-mass end of the IMF using deep VLT data and concluded that there is no substantial evidence for the effect of high stellar densities and the presence of numerous massive stars on the formation efficiency of low mass stars.

\subsection{Inference and interpretation of the universality of IMF}

We compare the results obtained in this study with similar studies across a broad range of star forming regions. On one hand we compare the results with nearby star forming regions in the solar neighbourhood whose low-mass and sub-stellar regimes of the IMF have been well constructed and on the other hand we compare with Galactic young massive clusters like Westerlund 1. Overlapping the log-normal  fit to the mass function of clusters in this study with a large number of open clusters and star forming regions in the solar neighborhood reveals that they are all consistent within the uncertainties over the same mass range (see \citealt{offner2014,moraux2016,suarez2019}).  We find that our results are in agreement with the general form of IMF given in \citet{bastian2010} stating that locally, there do not appear to be any strong systematic variation in the IMF. i.e, a log-normal distribution can well characterize the form of IMF within  0.08 - 3 $M_{\sun}$ for the clusters in our list. 

We conclude that there is no strong evidence for an effect that a combination of various  stellar densities, location in the Galaxy  and/or OB stellar radiation might have in the underlying form of  IMF or  characteristic mass among the clusters in this study and other resolved star forming regions and the Galactic field. For all star forming regions where star counts have been possible the stellar IMF appears to be very similar. The values of characteristic mass and $\sigma$ after assessing for the effect of various systematic uncertainties through Monte Carlo simulations are found to be close to the measured average IMF of all clusters within 3 sigma. We note that even if small variations between regions exist because of the environmental factors,  they may be hidden in the noise introduced from various  parameter estimation. 

Some of the uncertainties  present in this study are, the limited  completeness of the data towards the cluster center due to crowding, not accounting the multiplicity of various populations in each cluster, the effect of mass segregation and due to that some of the probable low-mass cluster members in the outer region might have not been included in the IMF analyses. However, with the existing data sets, we are unable to resolve these issues. Future high resolution imaging and spectroscopic analysis would help us resolve the above problems.

\section{Conclusions}

Beyond 500 pc from the Sun, in diverse environments at different Galactocentric radius where metallicity varies and massive stellar  feedback may dominate, the IMF has been measured only incompletely and imprecisely. The main goal of this work is to understand  the low-mass part of the IMF, and compare it with the well studied mass distributions in Galactic star forming regions. Since the young clusters ($<$ 5 Myr) are assumed to be less affected by dynamical evolution, their mass function can be considered as the IMF. We obtain  the  IMF of 8 young clusters (age $<$ 5Myr)  located at $\sim$ 1-4 kpc distance ($R_g$ $\sim$  6-12 kpc)  with a dense Pre-Main Sequence population and for the nearby cluster IC348 down to the brown dwarf regime. We use the deep near-IR data from United Kingdom Infrared Deep Sky Survey (UKIDSS) and Mayall Telescope at Kitt Peak National Observatory (KPNO) in J, H and K pass-bands along with Gaia DR2 data for the analysis. These clusters are embedded in massive stellar environments of radiation field strength $log(L_{FUV}/L_{\sun})$ $\sim$ 2.6 to 6.8,  $log(L_{EUV})$ $\sim$ 42.2 to 50.85  photons/s, with  stellar density in the range of $\sim$170 - 1220 stars/pc$^2$ and molecular cloud mass of $\sim$ 10$^4$ - 10$^5$ $M_{\sun}$. After a careful structural analysis, field star decontamination and completeness correction, we obtain an unbiased, uniformly sensitive sample of PMS members of the clusters down to brown-dwarf regime and obtain their form of IMF. The characteristic mass ($m_c$) and $\sigma$ values of these nine clusters lie in the range of $\sim$0.18 - 0.48 $M_\odot$ and $\sim$0.39 - 0.66 with a mean of 0.32$\pm$0.02 $M_\odot$ and 0.47$\pm$0.02, respectively. After accounting for the various sources of uncertainties through Monte Carlo based simulations, the mean values of  $m_c$ and $\sigma$ are estimated to be within 3 sigma uncertainties of above values. We compare the peak mass of IMF as well as its low-mass end with the nearby low-mass star-forming regions and with various super-massive clusters across the Milky Way to test the role of environmental factors. We also check for any systematic variation with respect to the radiation field strength, stellar density  as well as that of $R_g$. We conclude that there is no strong evidence for an effect that a combination of various stellar densities, location in the Galaxy and/or OB stellar radiation might have in the underlying form of IMF or characteristic mass among the clusters in this study. This work is the first of its kind to obtain the low-mass end of the IMF of a statistically rich sample of clusters using a unique method to verify the role of external factors on its universality.

\section*{Acknowledgements}

The authors are thankful to the referee for providing very constructive comments. This paper is based on data obtained as part of the UKIRT Infrared Deep Sky Survey. This publication made use of the data products from the Two Micron All Sky Survey (a joint project of the University of Massachusetts and the Infrared Processing and Analysis Center/California Institute of Technology, funded by NASA and NSF), archival data obtained with the Spitzer Space Telescope (operated by the Jet Propulsion Laboratory, California Institute of Technology, under a contract with NASA), the Wide-Field Infrared Survey Explorer (a joint project of the University of California, Los Angeles, and the Jet Propulsion Laboratory [JPL], California Institute of Technology [Caltech], funded by the National Aeronautics and Space Administration [NASA]), and the NOAO Science archive, which is operated by the Association of Universities for Research in Astronomy (AURA), Inc., under a cooperative agreement with the National Science Foundation. This publication makes use of VOSA, developed under the Spanish Virtual Observatory project supported by the Spanish MINECO through grant AyA2017-84089. VOSA has been partially updated by using funding from the European Union's Horizon 2020 Research and Innovation Programme, under Grant Agreement n$^\circ$ 776403 (EXOPLANETS-A). BD and JJ acknowledge the DST-SERB, Gov. of India  for the  start up research grant (No: SRG/2019/000664) for the financial support  and SP  acknowledges the DST-INSPIRE  fellowship  (No. IF180092) of the Department of Science \& Technology, India for carrying out this work. 

\section{Data Availability}

The UKIDSS and MYStIX datasets were derived from the public domain Vizier. The datasets for the W5 clusters will be available in the forthcoming paper. The data used to generate fig~\ref{fig:individual_imf} will be shared on reasonable request to the corresponding author.



\bibliographystyle{mnras}
\bibliography{main} 



\appendix

\section{Field of view of the clusters}
Figs. A1 and A2 are colour composite images using J, H, and K band UKIDSS images of all the sample clusters with the respective cluster radius marked with a green circle.
\label{sec:FOV_of_clusters}
\begin{figure*}
    \centering
    \includegraphics[scale=1.2, trim = 40 30 20 10 ]{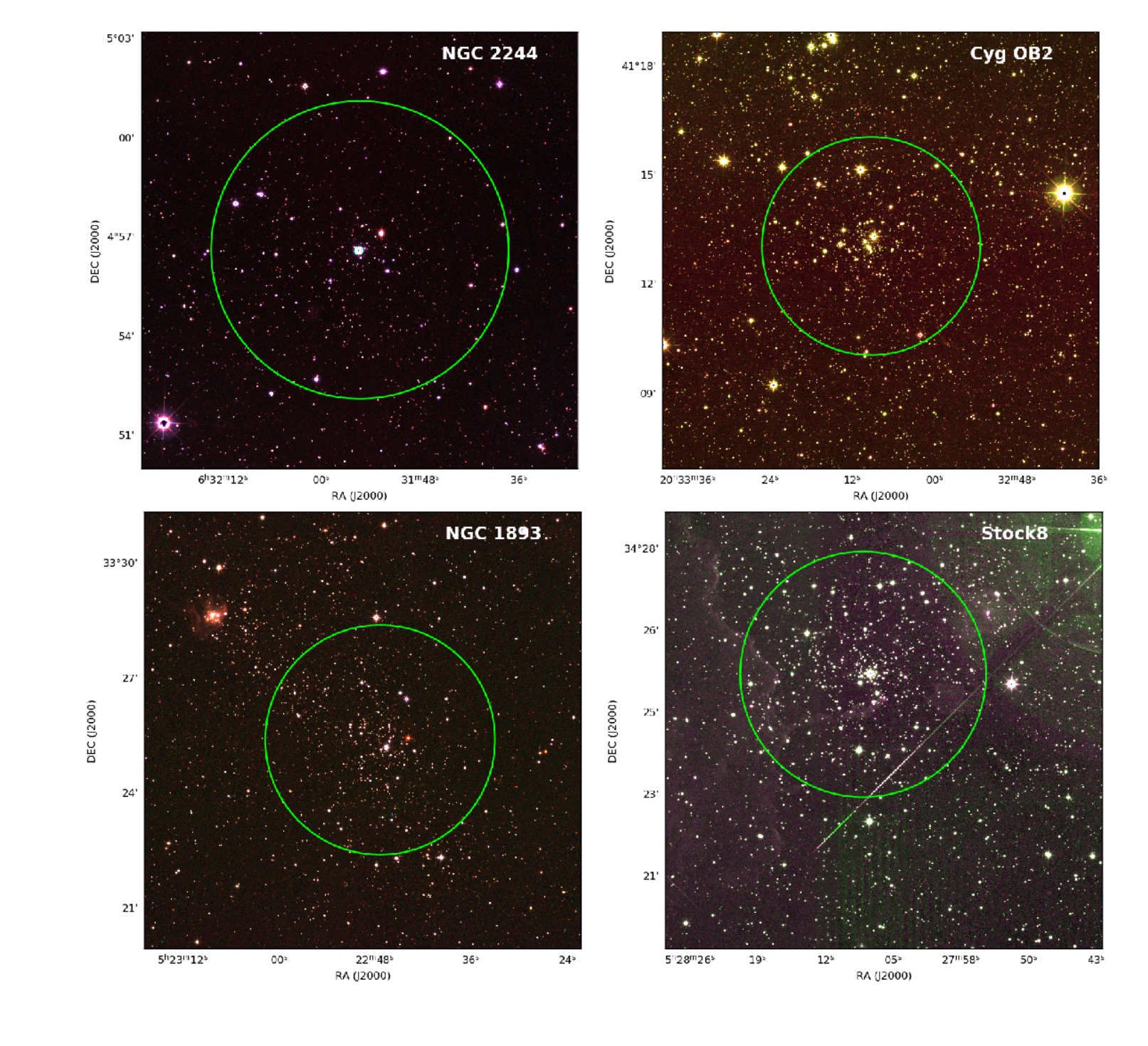}
    \caption{Color composite images using  J,H,K bands from UKIDSS for  NGC2244, Cygnus OB2, NGC1893 and Stock8. The green circle represents the radius of the respective cluster estimated in sec. \ref{sec:center&radius} (Refer table \ref{tab:phy_para})}
    \label{images1}
\end{figure*}

\begin{figure*}
    \centering
    \includegraphics[scale=1.2, trim = 40 30 20 10 ]{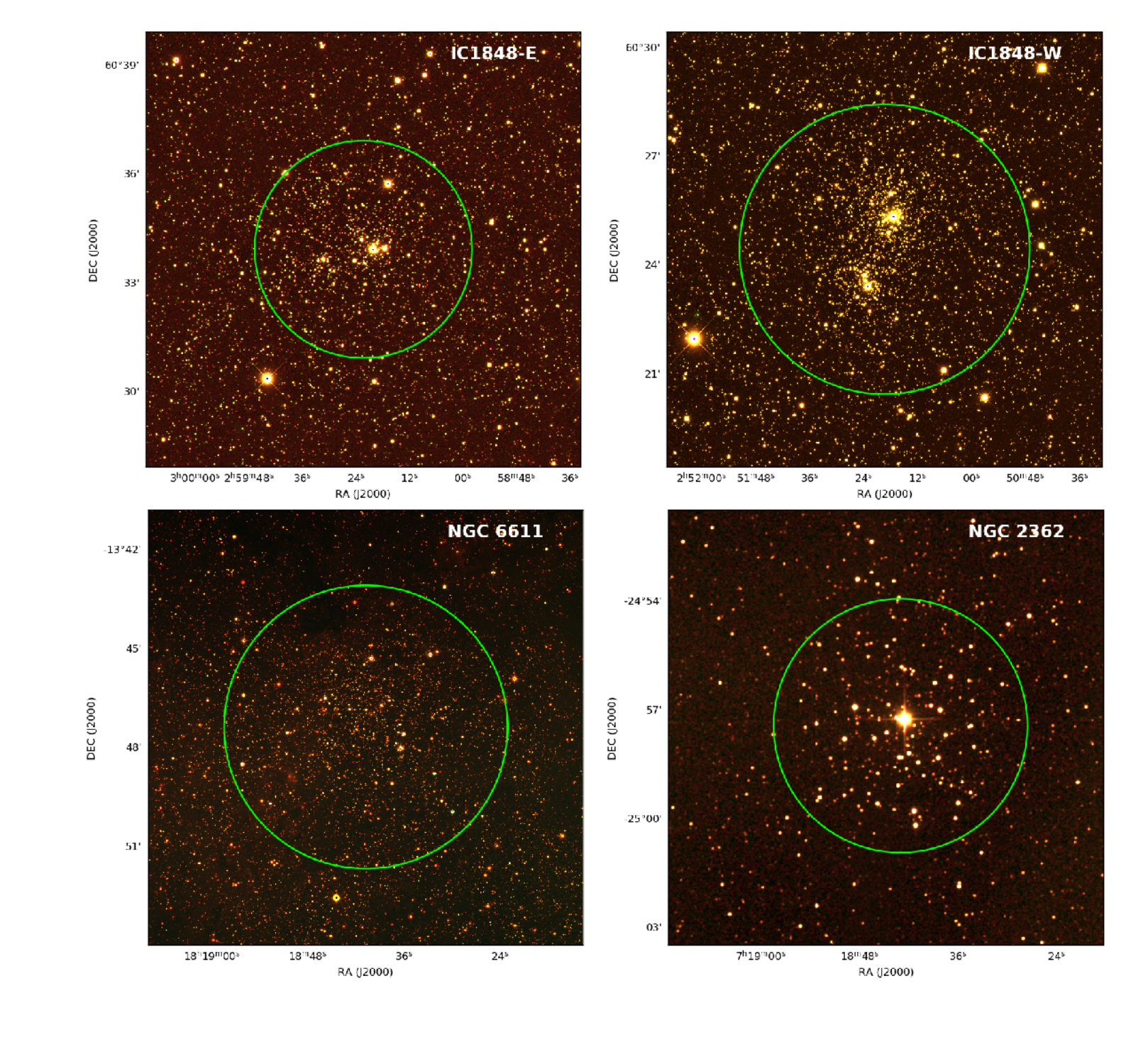}
    \caption{Color composite images using  J,H,K bands from  NEWFIRM for IC1848-E and  IC1848-W (top) and from UKIDSS for NGC6611. For NGC2362, the J,H,K images are taken from 2MASS archive as they were unavailable in UKIDSS archive.  The green circle represents the radius of the respective cluster estimated in sec. \ref{sec:center&radius} (Refer table \ref{tab:phy_para})}
    \label{images2}
\end{figure*}

\section{Completeness histograms of a few control fields}
The completeness histograms of some of the control fields of the sample clusters in J and K bands are shown in Fig. B1.
\label{sec:cf_compl_hist}
\begin{figure*}
    \centering
    \includegraphics[width=\textwidth]{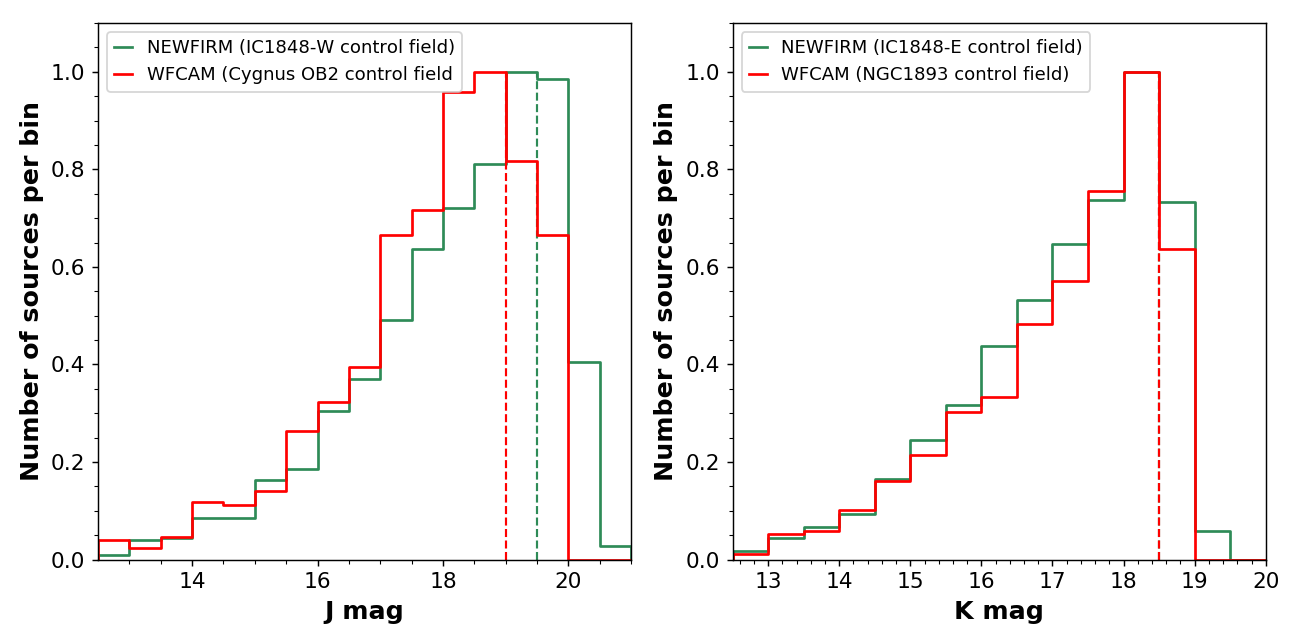}
    \caption{Normalized sample histograms showing the completeness limits of the control fields corresponding to the clusters in Fig~\ref{fig:completeness_hist}. The turnover point in the distribution serves as a proxy for the completeness limit of the data. The dashed lines mark the $\sim$ $90\%$ completeness limit of the photometry.}
    \label{fig:cf_completeness_hist}
\end{figure*}

\section{Cluster, control field and field  decontaminated CMDs}
The field star decontamination process using colour-magnitude diagram is depicted in Figs. C1-C8.
\label{sec:fd_CMD_of_all_clusters}

\begin{figure*}
    \centering
    \includegraphics[width=\textwidth]{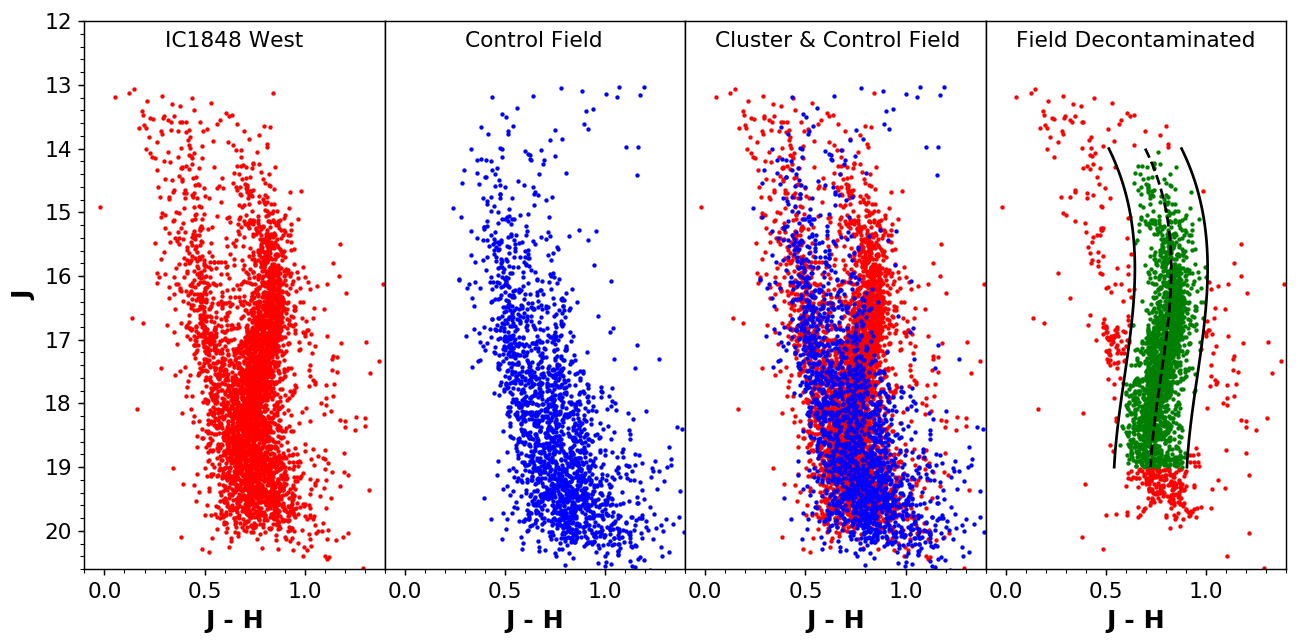}
    \caption{CMDs  of IC1848-W.}
    \label{fig:fd_ic1848w}
\end{figure*}

\begin{figure*}
    \centering
    \includegraphics[width=\textwidth]{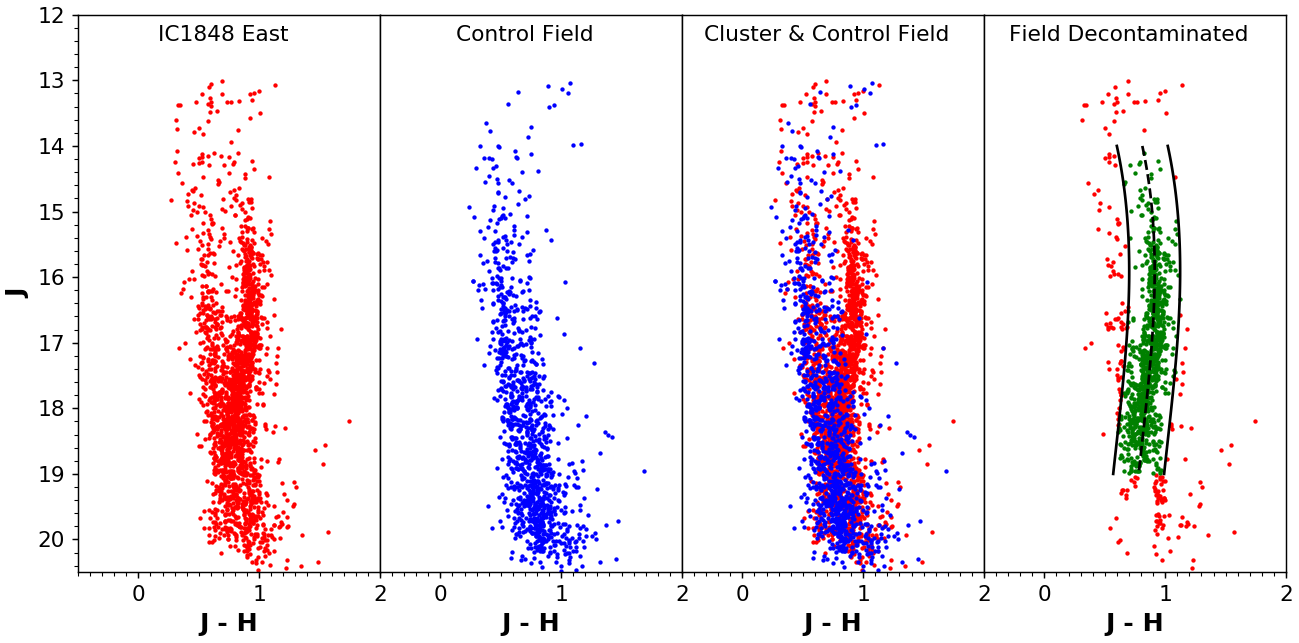}
    \caption{CMDs  of IC1848-E.}
    \label{fig:fd_ic1848e}
\end{figure*}

\begin{figure*}
    \centering
    \includegraphics[width=\textwidth]{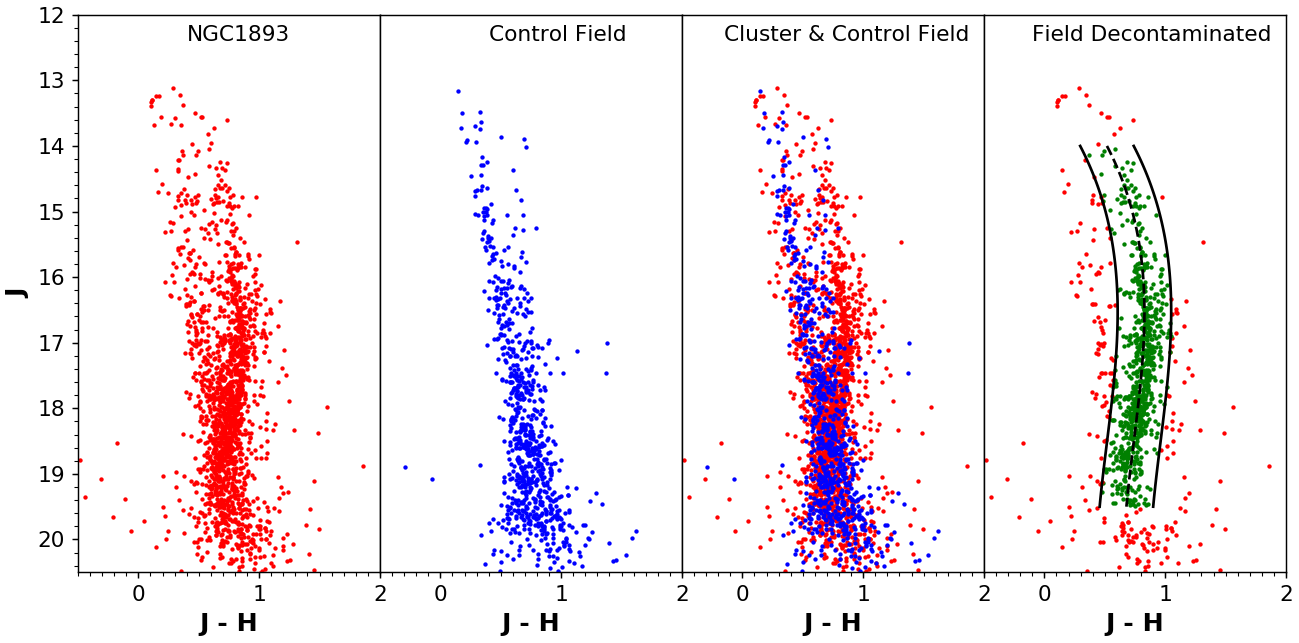}
    \caption{CMDs of NGC1893.}
    \label{fig:fd_ngc1893}
\end{figure*}

\begin{figure*}
    \centering
    \includegraphics[width=\textwidth]{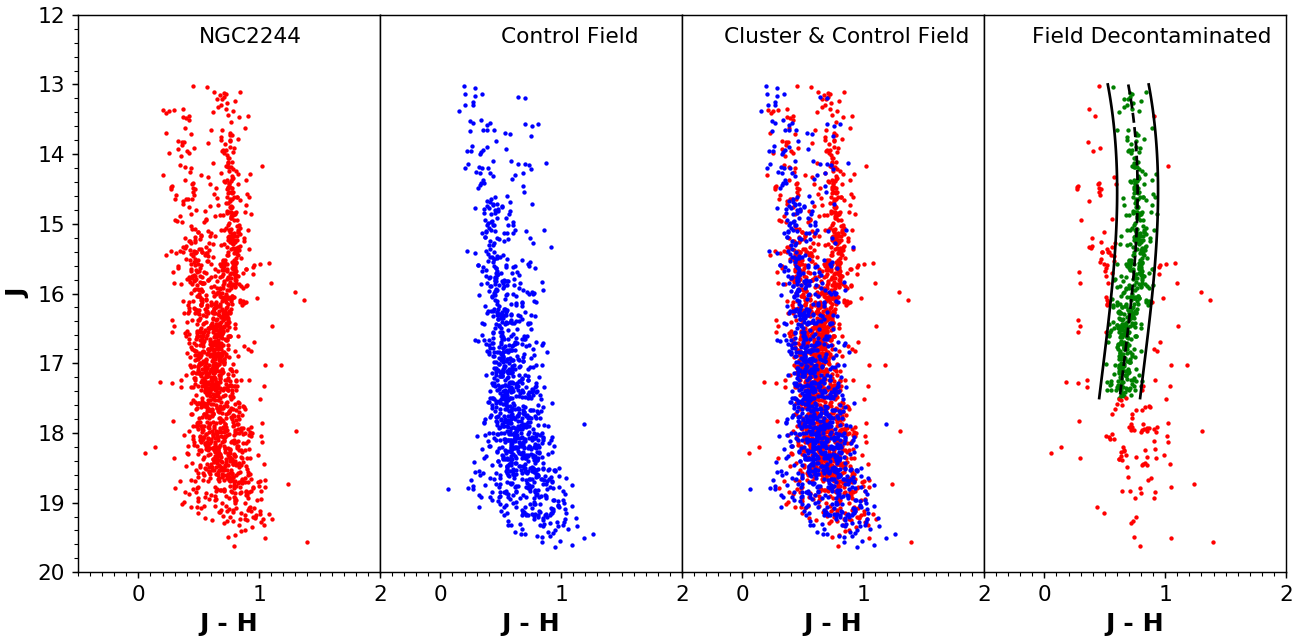}
    \caption{CMDs of NGC2244.}
    \label{fig:fd_ngc2244}
\end{figure*}

\begin{figure*}
    \centering
    \includegraphics[width=\textwidth]{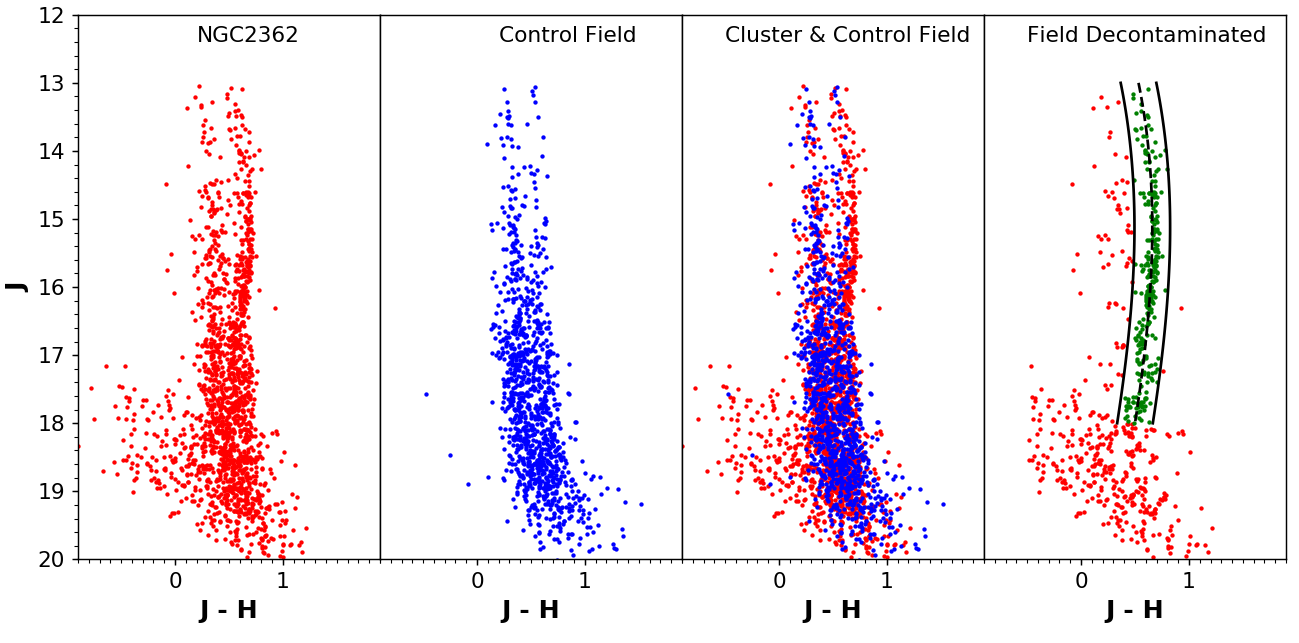}
    \caption{CMDs of NGC2362.}
    \label{fig:fd_ngc2362}
\end{figure*}

\begin{figure*}
    \centering
    \includegraphics[width=\textwidth]{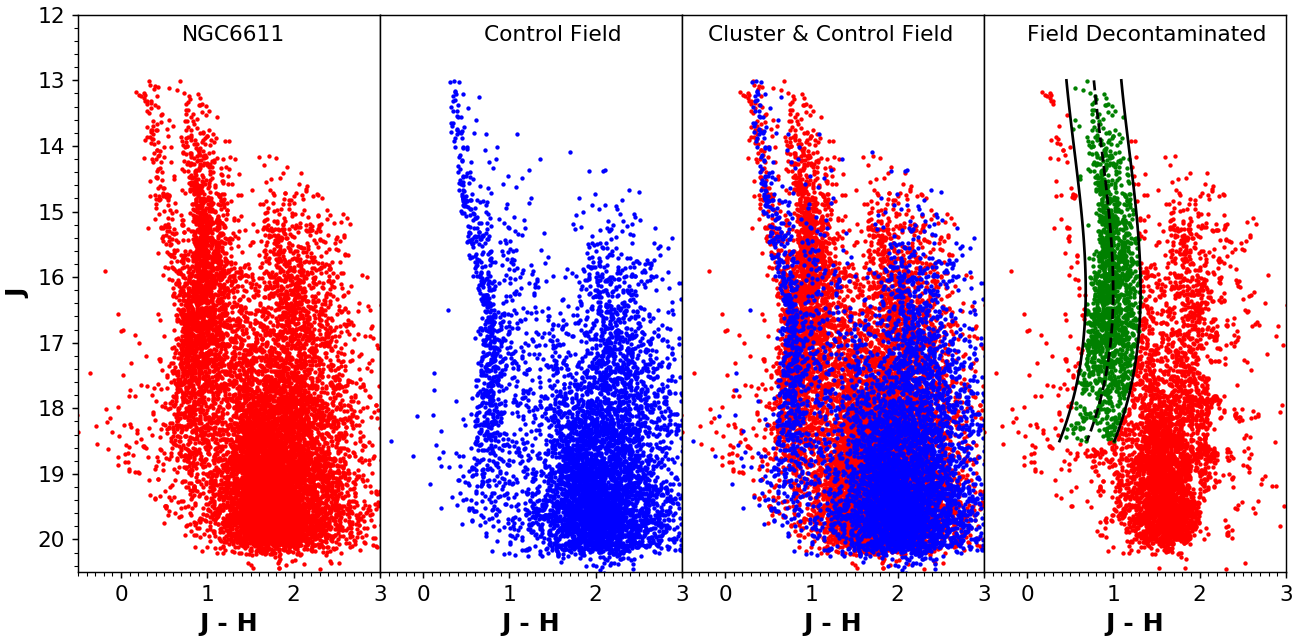}
    \caption{CMDs of NGC6611.}
    \label{fig:fd_ngc6611}
\end{figure*}

\begin{figure*}
    \centering
    \includegraphics[width=\textwidth]{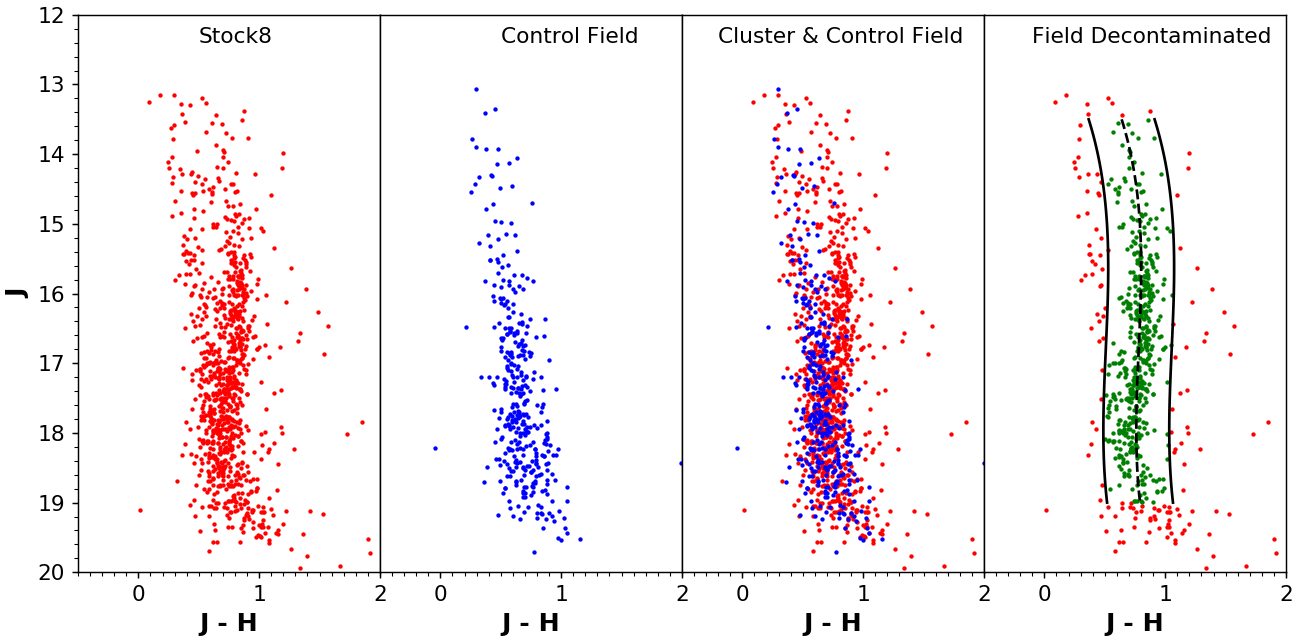}
    \caption{CMDs of Stock8.}
    \label{fig:fd_stock8}
\end{figure*}

\begin{figure*}
    \centering
    \includegraphics[width=\textwidth]{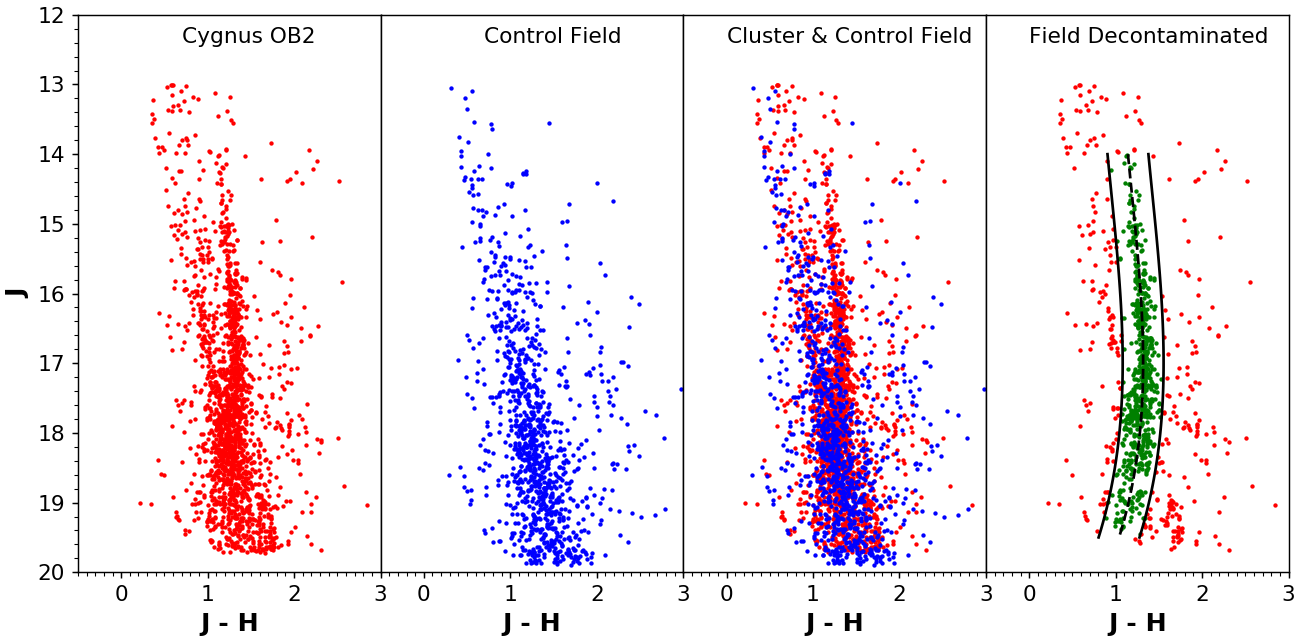} 
    \caption{CMDs of Cygnus OB2.}
    \label{fig:fd_cygnusOB2}
\end{figure*}


\bsp	
\label{lastpage}
\end{document}